\documentclass[11pt]{article}
\pdfoutput=1
\usepackage{graphicx}
\usepackage{float}
\usepackage{epsfig}
\usepackage{ulem}
\usepackage{latexsym,amsmath,amsfonts,amssymb}
\usepackage[latin1]{inputenc}
\usepackage{rotating}
\usepackage[american]{babel}
\usepackage{bbm}
\usepackage{color}
\usepackage{slashed}
\usepackage[unicode]{hyperref}
\usepackage{lscape}
\def\im{Invent. Math.}

\def\a{\alpha}
\def\b{\beta}
\def\c{\gamma}
\def\d{\delta}
\def\f{\phi}               
\def\vf{\varphi}  
\def\tvf{\tilde{\varphi}}
\def\vp{\varphi}
\def\g{\gamma}
\def\h{\eta}
\def\j{\psi}
\def\k{\kappa}                    
\def\l{\lambda}
\def\m{\mu}
\def\n{\nu}
\def\o{\omega}  \def\w{\omega}

\def\q{\theta}  \def\th{\theta}                  
\def\r{\rho}                                     
\def\s{\sigma}                                   
\def\t{\tau}
\def\u{\upsilon}
\def\x{\xi}
\def\z{\zeta}
\def\pt{\tilde{\varphi}}
\def\tt{\tilde{\theta}}
\def\lab{\label}
\def\6{\partial}
\def\wg{\wedge}
\def\bpsi{\bar{\psi}}
\def\bt{\bar{\theta}}
\def\bvf{\bar{\varphi}}

\DeclareMathOperator{\tr}{tr}

\newcommand{\be}{\begin{equation}}
\newcommand{\ee}{\end{equation}}
\newcommand{\beq}{\begin{equation}}
\newcommand{\eeq}{\end{equation}}
\newcommand{\bea}{\begin{eqnarray}}
\newcommand{\eea}{\end{eqnarray}}
\newcommand{\nn}{\nonumber}

\newcommand{\ba}{\begin{eqnarray}}
\newcommand{\ea}{\end{eqnarray}}

\newcommand{\beqs}{\begin{eqnarray}}
\newcommand{\eeqs}{\end{eqnarray}}
\newcommand{\bal}{\begin{aligned}}
\newcommand{\eal}{\end{aligned}}

\begin{document}
\baselineskip=15.5pt
\pagestyle{plain}
\setcounter{page}{1}

\def\del{{\partial}}
\def\vev#1{\left\langle #1 \right\rangle}
\def\cn{{\cal N}}
\def\co{{\cal O}}


\def\IC{{\mathbb C}}
\def\IR{{\mathbb R}}
\def\IZ{{\mathbb Z}}
\def\RP{{\bf RP}}
\def\CP{{\bf CP}}
\def\Poincare{{Poincar\'e }}
\def\tr{{\rm tr}}
\def\tp{{\tilde \Phi}}

\def\TL{\hfil$\displaystyle{##}$}
\def\TR{$\displaystyle{{}##}$\hfil}
\def\TC{\hfil$\displaystyle{##}$\hfil}
\def\TT{\hbox{##}}
\def\HLINE{\noalign{\vskip1\jot}\hline\noalign{\vskip1\jot}}
\def\seqalign#1#2{\vcenter{\openup1\jot
   \halign{\strut #1\cr #2 \cr}}}
\def\lbldef#1#2{\expandafter\gdef\csname #1\endcsname {#2}}
\def\eqn#1#2{\lbldef{#1}{(\ref{#1})}%
\begin{equation} #2 \label{#1} \end{equation}}
\def\eqalign#1{\vcenter{\openup1\jot
     \halign{\strut\span\TL & \span\TR\cr #1 \cr
    }}}

\def\eno#1{(\ref{#1})}
\def\href#1#2{#2}
\def\half{\frac{1}{2}}



\def\ads{{\it AdS}}
\def\adsp{{\it AdS}$_{p+2}$}
\def\cft{{\it CFT}}

\newcommand{\ber}{\begin{eqnarray}}
\newcommand{\eer}{\end{eqnarray}}

\newcommand{\beqar}{\begin{eqnarray}}
\newcommand{\cN}{{\cal N}}
\newcommand{\cO}{{\cal O}}
\newcommand{\cA}{{\cal A}}
\newcommand{\cT}{{\cal T}}
\newcommand{\cF}{{\cal F}}
\newcommand{\cC}{{\cal C}}
\newcommand{\cR}{{\cal R}}
\newcommand{\cW}{{\cal W}}
\newcommand{\eeqar}{\end{eqnarray}}
\newcommand{\tht}{\thteta}
\newcommand{\lm}{\lambda}\newcommand{\Lm}{\Lambda}


\newcommand{\nonu}{\nonumber}
\newcommand{\oh}{\displaystyle{\frac{1}{2}}}
\newcommand{\dsl}
   {\kern.06em\hbox{\raise.15ex\hbox{$/$}\kern-.56em\hbox{$\partial$}}}
\newcommand{\id}{i\!\!\not\!\partial}
\newcommand{\as}{\not\!\! A}
\newcommand{\ps}{\not\! p}
\newcommand{\ks}{\not\! k}
\newcommand{\D}{{\cal{D}}}
\newcommand{\dv}{d^2x}
\newcommand{\Z}{{\cal Z}}
\newcommand{\N}{{\cal N}}
\newcommand{\Dsl}{\not\!\! D}
\newcommand{\Bsl}{\not\!\! B}
\newcommand{\Psl}{\not\!\! P}

\newcommand{\eeqarr}{\end{eqnarray}}
\newcommand{\ZZ}{{\rm \kern 0.275em Z \kern -0.92em Z}\;}


\def\del{{\delta^{\hbox{\sevenrm B}}}} \def\ex{{\hbox{\rm e}}}
\def\azb{A_{\bar z}} \def\az{A_z} \def\bzb{B_{\bar z}} \def\bz{B_z}
\def\czb{C_{\bar z}} \def\cz{C_z} \def\dzb{D_{\bar z}} \def\dz{D_z}
\def\im{{\hbox{\rm Im}}} \def\mod{{\hbox{\rm mod}}} \def\tr{{\hbox{\rm Tr}}}
\def\ch{{\hbox{\rm ch}}} \def\imp{{\hbox{\sevenrm Im}}}
\def\trp{{\hbox{\sevenrm Tr}}} \def\vol{{\hbox{\rm Vol}}}
\def\rl{\Lambda_{\hbox{\sevenrm R}}} \def\wl{\Lambda_{\hbox{\sevenrm W}}}
\def\fc{{\cal F}_{k+\cox}} \def\vev{vacuum expectation value}
\def\nodiv{\mid{\hbox{\hskip-7.8pt/}}}
\def\ie{{\em i.e.}}
\def\ie{\hbox{\it i.e.}}

\def\CC{{\mathchoice
{\rm C\mkern-8mu\vrule height1.45ex depth-.05ex
width.05em\mkern9mu\kern-.05em}
{\rm C\mkern-8mu\vrule height1.45ex depth-.05ex
width.05em\mkern9mu\kern-.05em}
{\rm C\mkern-8mu\vrule height1ex depth-.07ex
width.035em\mkern9mu\kern-.035em}
{\rm C\mkern-8mu\vrule height.65ex depth-.1ex
width.025em\mkern8mu\kern-.025em}}}

\def\RR{{\rm I\kern-1.6pt {\rm R}}}
\def\NN{{\rm I\!N}}
\def\ZZ{{\rm Z}\kern-3.8pt {\rm Z} \kern2pt}
\def\IB{\relax{\rm I\kern-.18em B}}
\def\ID{\relax{\rm I\kern-.18em D}}
\def\II{\relax{\rm I\kern-.18em I}}
\def\IP{\relax{\rm I\kern-.18em P}}
\newcommand{\CS}{{\scriptstyle {\rm CS}}}
\newcommand{\CSs}{{\scriptscriptstyle {\rm CS}}}
\newcommand{\rc}{\nonumber\\}
\newcommand{\bear}{\begin{eqnarray}}
\newcommand{\eear}{\end{eqnarray}}

\newcommand{\LL}{{\cal L}}

\def\mani{{\cal M}}
\def\calo{{\cal O}}
\def\calb{{\cal B}}
\def\calw{{\cal W}}
\def\calz{{\cal Z}}
\def\cald{{\cal D}}
\def\calc{{\cal C}}

\def\to{\rightarrow}
\def\ele{{\hbox{\sevenrm L}}}
\def\ere{{\hbox{\sevenrm R}}}
\def\zb{{\bar z}}
\def\wb{{\bar w}}
\def\nodiv{\mid{\hbox{\hskip-7.8pt/}}}
\def\menos{\hbox{\hskip-2.9pt}}
\def\dr{\dot R_}
\def\drr{\dot r_}
\def\ds{\dot s_}
\def\da{\dot A_}
\def\dga{\dot \gamma_}
\def\ga{\gamma_}
\def\dal{\dot\alpha_}
\def\al{\alpha_}
\def\cl{{closed}}
\def\cls{{closing}}
\def\vev{vacuum expectation value}
\def\tr{{\rm Tr}}
\def\to{\rightarrow}
\def\too{\longrightarrow}


\def\a{\alpha}
\def\b{\beta}
\def\c{\gamma}
\def\d{\delta}
\def\e{\epsilon}           
\def\F{\Phi}
\def\f{\phi}               
\def\vf{\varphi}  \def\tvf{\tilde{\varphi}}
\def\vp{\varphi}
\def\g{\gamma}
\def\h{\eta}
\def\j{\psi}
\def\k{\kappa}                    
\def\l{\lambda}
\def\m{\mu}
\def\n{\nu}
\def\o{\omega}  \def\w{\omega}
\def\q{\theta}  \def\th{\theta}                  
\def\r{\rho}                                     
\def\s{\sigma}                                   
\def\t{\tau}
\def\u{\upsilon}
\def\x{\xi}
\def\X{\Xi}
\def\z{\zeta}
\def\pt{\tilde{\varphi}}
\def\tt{\tilde{\theta}}
\def\lab{\label}
\def\6{\partial}
\def\wg{\wedge}
\def\atanh{{\rm arctanh}}
\def\bpsi{\bar{\psi}}
\def\bt{\bar{\theta}}
\def\bvf{\bar{\varphi}}

%



\newfont{\namefont}{cmr10}
\newfont{\addfont}{cmti7 scaled 1440}
\newfont{\boldmathfont}{cmbx10}
\newfont{\headfontb}{cmbx10 scaled 1728}





\newcommand{\re}{\,\mathbb{R}\mbox{e}\,}
\newcommand{\hyph}[1]{$#1$\nobreakdash-\hspace{0pt}}
\providecommand{\abs}[1]{\lvert#1\rvert}
\newcommand{\Nugual}[1]{$\mathcal{N}= #1 $}
\newcommand{\sub}[2]{#1_\text{#2}}
\newcommand{\partfrac}[2]{\frac{\partial #1}{\partial #2}}
\newcommand{\bsp}[1]{\begin{equation} \begin{split} #1 \end{split} \end{equation}}
\newcommand{\calF}{\mathcal{F}}
\newcommand{\calO}{\mathcal{O}}
\newcommand{\calM}{\mathcal{M}}
\newcommand{\calV}{\mathcal{V}}
\newcommand{\bbZ}{\mathbb{Z}}
\newcommand{\bbC}{\mathbb{C}}
\newcommand{\cK}{{\cal K}}

\newcommand{\Thq}{\Theta\left(\r-\r_q\right)}
\newcommand{\Dq}{\d\left(\r-\r_q\right)}
\newcommand{\kten}{\kappa^2_{\left(10\right)}}
\newcommand{\pbi}[1]{\imath^*\left(#1\right)}
\newcommand{\ho}{\hat{\omega}}
\newcommand{\tth}{\tilde{\th}}
\newcommand{\tf}{\tilde{\f}}
\newcommand{\tj}{\tilde{\j}}
\newcommand{\tw}{\tilde{\omega}}
\newcommand{\tz}{\tilde{z}}
\newcommand{\prj}[2]{(\partial_r{#1})(\partial_{\j}{#2})-(\partial_r{#2})(\partial_{\j}{#1})}
\def\atanh{{\rm arctanh}}
\def\sech{{\rm sech}}
\def\csch{{\rm csch}}
\allowdisplaybreaks[1]

\def\red{\textcolor[rgb]{0.98,0.00,0.00}}

\newcommand{\Dan}[1] {{\textcolor{blue}{#1}}}

\numberwithin{equation}{section}

\newcommand{\Tr}{\mbox{Tr}}    


%

\setcounter{footnote}{0}
\renewcommand{\theequation}{{\rm\thesection.\arabic{equation}}}

\begin{titlepage}

\vspace{0.1in}

\begin{center}
\Large \bf {Three-dimensional $\mathcal{N}=4$ Linear Quivers and non-Abelian T-duals}
\end{center}

\vskip 0.2truein
\begin{center} 
{\bf Yolanda Lozano}$^{a,}$\footnote{ylozano@uniovi.es}, {\bf Niall T. Macpherson}$^{b,}$\footnote{niall.macpherson@mib.infn.it}, {\bf Jes\'us Montero}$^{a,}$\footnote{monteroaragon@uniovi.es} and {\bf Carlos N\'u\~nez}$^{c,}$\footnote{c.nunez@swansea.ac.uk}
\end{center}

\begin{center}
\vspace{.2in}
 $a$: Department of Physics, University of Oviedo,
Avda. Calvo Sotelo 18, 33007 Oviedo, Spain
\vskip 3mm
 $b$: Dipartimento di Fisica, Universit\`a di Milano--Bicocca, I-20126 Milano, Italy\\
 and \\
 INFN, sezione di Milano--Bicocca
\vskip 3mm
 $c$: Department of Physics, Swansea University, Swansea SA2 8PP, United Kingdom.
\end{center}
\vskip 10mm

\centerline{\bf Abstract:}
 In this paper we construct a new Type IIB background with an $AdS_4$ factor that preserves ${\cal N}=4$ Supersymmetry. This  solution is obtained using a non-Abelian T-duality transformation
on the Type IIA reduction of the $AdS_4\times S^7$ background. We interpret our configuration  as a patch of a more general background with localised sources, dual to the renormalisation fixed point of a  
 $T_{\rho}^{\hat{\rho}} (SU(N))$ quiver field theory. This relates explicitly the $AdS_4$ geometry to a D3-D5-NS5 brane intersection, illuminating what seems to be a more general phenomenon, relating $AdS_{p+1}$ backgrounds generated by non-Abelian T-duality to Dp-D(p+2)-NS5 branes intersections.

\smallskip
\end{titlepage}
\setcounter{footnote}{0}

\tableofcontents

\setcounter{footnote}{0}
\renewcommand{\theequation}{{\rm\thesection.\arabic{equation}}}

\section{Introduction }

The idea of duality is very old, perhaps dating back to the (self) duality 
of the Maxwell equations in the absence of charges and currents.
The transformation of the fields describing a given dynamics into a 
{\it different} set of fields where particular phenomena become more
apparent, is a recurrent idea in Theoretical Physics.
Indeed, dualities like those proposed by Montonen and Olive
\cite{Montonen:1977sn},
Seiberg and Witten \cite{Seiberg:1994rs},
Seiberg \cite{Seiberg:1994pq},
or the U-duality web in String Theory (see for example \cite{Sen:1998kr}) 
are examples of this. While these dualities are very hard to prove (hence initially conjectured), they have
very far reaching consequences in Physics: 
the phenomena that in one description are
highly fluctuating and hence eminently quantum mechanical, 
become semiclassical and characteristically 
weakly coupled in the dual set of variables.
The AdS/CFT duality 
\cite{Maldacena:1997re}
relating gauge theories and String theories is a paradigmatic
example of this.

Other dualities, like the Kramers-Wannier self duality of the two-dimensional
Ising model \cite{Kramers:1941kn},
bosonisation in two dimensions \cite{Coleman:1974bu}
or T-duality in the String Theory sigma model \cite{Buscher:1987qj,Rocek:1991ps},
are within the class of dualities that can be formally proven.

In 1993, Quevedo and de la Ossa \cite{delaOssa:1992vci},
 following ideas in 
\cite{Rocek:1991ps},
proposed a non-Abelian generalisation of T-duality, applicable to the Neveu-Schwarz sector of the string sigma model. This was later complemented
by Sfetsos and Thompson, who showed how to
transform the fields in the Ramond sector \cite{Sfetsos:2010uq}.
This important work  opened the way for further study involving new
backgrounds and illuminating some geometrical and dual field theoretic aspects of the  non-Abelian T-duality
 \cite{varios1}-\cite{Lozano:2016kum}. These works have in turn motivated the search for new classes of supersymetric $AdS$ solutions that were overlooked until recently \cite{Apruzzi:2014qva}-\cite{Couzens:2016iot}.

Whilst the sigma-model
procedure to calculate the non-Abelian T-dual of a given
background is apparently straightforward, 
many interesting subtleties related to
global aspects and invertibility of the duality arise. 
These subtle aspects were studied in the mid-nineties
but not completely resolved, in spite of many serious attempts
\cite{Alvarez:1993qi}-\cite{Klimcik:1995ux}.
Some of such concrete problems are the (im)possibility of extending the non-Abelian duality procedure to all orders in string perturbation theory and $\alpha'$, and
the determination of the range of
the coordinates and topology of the dual manifold. These issues cast doubts about the 'duality-character' of the non-Abelian T-duality
transformation.

One goal of this paper --elaborating on ideas introduced in
\cite{Lozano:2016kum}--
is to get information on some of the global problems mentioned above. 
The example we will
consider here involves a Type IIB background with an $AdS_4$
factor, preserving ${\cal N}=4$ Supersymmetry.

A second goal of this paper --of interest in a broader context-- will be to produce a new analytic solution
to the Type IIB Supergravity equations of motion with an $AdS_4$ factor, that can be interpreted as an intersection of D3-D5-NS5 branes. Our example illuminates what is
surely a more general phenomenon, relating $AdS_{p+1}$ geometries generated by non-Abelian T-duality with $Dp$-$D{(p+2)}$-NS5 branes intersections---see for example
\cite{Apruzzi:2014qva,Gaiotto:2014lca} for other recent studies of such configurations.

Furthermore, our case-study provides an interesting arena where the CFT interpretation of non-Abelian T-duality put forward in \cite{Lozano:2016kum} can be tested. Indeed, using the results in 
\cite{Assel:2011xz,Assel:2012cj} (see also \cite{Aharony:2011yc}), which elaborate on certain limits of Type IIB Supergravity solutions discussed in \cite{D'Hoker:2007xz,D'Hoker:2007xy}, it is possible to associate a concrete CFT dual to our $AdS_4$ solution. This will be a
${\cal N}=4$, d=3 conformal field theory, arising as the Renormalisation Group fixed point of 
a $T_{{\rho}}^{\hat{\rho}}(SU(N))$ quantum field theory that belongs to the general class
introduced by Gaiotto and Witten in \cite{Gaiotto:2008ak}.
These conformal field theories can be described in terms of  a linear quiver with bi-fundamental and fundamental matter
or, equivalently,  in terms of Hanany-Witten set ups \cite{Hanany:1996ie} containing D3, NS5 and D5 branes.



This work extends the ideas in 
\cite{Lozano:2016kum} to the $AdS_4/CFT_3$ case.
The paper
\cite{Lozano:2016kum}
deals with the {\it singular} background
obtained by the application of
non-Abelian T-duality on $AdS_5\times S^5$ and its interpretation
as a Gaiotto-Maldacena type of geometry
\cite{Gaiotto:2009gz}.
Using the formal developments of
 \cite{ReidEdwards:2010qs}, 
a {\it completion} to the geometry generated by non-Abelian duality was proposed, with the following relevant properties:
\begin{itemize}
\item{It is a smooth background, except at isolated points where  brane sources are located.}
\item{The dual CFT is known explicitly. }
\item{The coordinates of the completed geometry have a definite range, determined by imposing 
the matching between observables calculated with
 the CFT and with  the geometrical description.}
\item{The original non-Abelian T-dual  background (that is, the geometry before {\it completion}) can be seen as a {\it zoom-in}
on a patch of the completed manifold.
  }
\end{itemize}
In this paper, we will use a combination of insights from three-dimensional ${\cal N}=4$ CFTs
and their dual geometries to obtain a similar understanding
of an $AdS_4$ Type IIB background, obtained by the action of non-Abelian T-duality on the Type IIA
reduction of $AdS_4\times S^7$. An outline of this works goes as follows.

In Section \ref{sec:sugra_solutions}, we present our (new) background, analyse the amount of
SUSY preserved and study the structure of its singularities.
The calculation of the associated charges leads us 
to a proposal for the Hanany-Witten set-up \cite{Hanany:1996ie}.
In Section \ref{seccion3}
we discuss aspects of  ${\cal N}=4$ SCFTs in three dimensions. The associated
backgrounds containing an $AdS_4$ sub-manifold are also discussed.
In Sections \ref{seccion5} and \ref{seccion5bis}, we embed our non-abelian T-dual geometry into the formalism of \cite{Assel:2011xz} (ABEG hereafter).
This leads us to a precise proposal for the CFT dual to our background. We
interpret our singular solution as embedded in a more generic background
(with the characteristics itemized above).
Section \ref{seccion6} discusses the subtle calculation of the free energy for the
CFT defined by the non-abelian T-dual geometry.
Conclusions and some further directions to explore are collected in Section \ref{seccionconclusiones}.  Appendix \ref{section:ATD} summarises the main properties of the Abelian T-dual limit of the non-Abelian solution, of relevance for the interpretation of the free energy. Finally, Appendix B contains an interesting general relation between Abelian and non-Abelian T-duals.

\section{The Type IIB $\mathcal{N}=4$ $AdS_4$ solution}\label{sec:sugra_solutions}

In this section we present the new type IIB $\mathcal{N}=4$ $AdS_4$ background  where our ideas will be tested. It is generated from the maximally 
supersymmetric $AdS_4\times S^7$ solution in M-theory (once reduced to Type IIA), through a non-Abelian T-duality transformation.

To begin we parametrise the M-theory solution such that we manifestly have two three-spheres $S^3_1$ and $S^3_2$, as
\begin{align}\label{eq: ads4s7}
ds^2_{11d}&= ds^2(AdS_4)+4L^2\bigg(\frac{1}{4}d\mu^2+ \sin^2\left(\frac{\mu}{2}\right) ds^2\left(S_1^3\right)+\cos^2\left(\frac{\mu}{2}\right) ds^2\left(S_2^3\right)\bigg),\nn\\[2mm]
G_4 &= \frac{3\rho^2}{L^3}dt\wedge dx_1\wedge dx_2\wedge d\rho= \frac{3}{L}\text{Vol}(AdS_4),~~~~ds^2(AdS_4)=\frac{\rho^2}{L^2}dx_{1,2}^2 +L^2\frac{d\rho^2}{\rho^2},
\end{align}
where as usual for $AdS_4$ Freund-Rubin solutions the $AdS$ and internal radii obey the relation $R_{S^7}=2 R_{AdS_4}$. We take the three-spheres to have unit radius, which means $\mu\in[0,\pi)$. With the above parametrisation there is enough symmetry to reduce to IIA within one of the three spheres and then perform a T-duality transformation on the other. Here we will focus on performing an $SU(2)$ non-Abelian T-duality on the residual $SU(2)$. We also give details of the Hopf fibre T-dual in  Appendix \ref{section:ATD}.

We want to reduce to Type IIA on the Hopf direction of $S^3_2$ by parametrising it as
\beq
4ds(S^3_2)= d\theta_2^2+ \sin^2\theta_2 d\phi_2^2+ (d\psi_2+\cos\theta_2 d\phi_2)^2,
\eeq
with $\psi_2\in [0,4\pi]$. Since some supersymmetry will be broken in the process, 
as the isometry parametrised by $\partial_{\psi_2}$ defines a $U(1)$ subgroup of the full $SO(8)$ R-symmetry,
we briefly study the Killing spinor equations. To this end we introduce the manifestly $U(1)_{\psi_2}$ invariant vielbein,
\bea\label{eq: M-theoryFrame}
& & e^{xi}= \frac{R}{L} dx_i~~(i=t,x_1,x_2), \;\; e^R= \frac{L}{R} dR,\nonumber\\
& & e^\mu= L d\mu,\;\; e^{1}= L \sin(\frac{\mu}{2})\omega_1,\;\; e^{2}=L\sin(\frac{\mu}{2})\omega_2,\;\; e^{3}=L\sin(\frac{\mu}{2})\omega_3.\nonumber\\
& &  e^{\theta_2}= L \cos(\frac{\mu}{2})d\theta_2,\;\; e^{\phi_2}=L\cos(\frac{\mu}{2}) \sin\theta_2 d\phi_2,\;\; e^{\psi_2}=L\cos(\frac{\mu}{2})(d\psi_2 +\cos\theta_2 d\phi_2).\nonumber\\
& &  F_4=\frac{3}{L}e^{t x_1 x_2 \rho},\label{11dconfiguration}
\eea
where 
\beq
\omega_1+ i \omega_2 =e^{i\psi_1}\big(i d\theta_1+\sin\theta_1 d\phi_1\big),~~~ \omega_3= d\psi_1+ \cos\theta_1 d\phi_1,
\eeq
which makes manifest an additional $SU(2)$ isometry parametrised by $S^3_{1}$.
The gravitino variation on $S^7$ is given in flat indices by \footnote{That the $AdS_4$ directions solve is a standard exercise that we omit for brevity.}
\beq
\nabla_a \eta+ \frac{1}{4 L}\Gamma_a\hat{\gamma}\eta=0,
\eeq
where $\hat{\gamma}=\Gamma_t\Gamma_{x_1}\Gamma_{x_2}\Gamma_{\rho}$. The number of preserved supercharges is determined by what fraction of the initial 32 SUSYs are consistent with setting $\partial_{\psi_2}\eta =0$ in the frame of eq. \eqref{eq: M-theoryFrame}. One can show by imposing that $\eta$ is independent of $\psi_2$ that one is lead to a single projection that the Killing spinor must obey,
\beq
\Gamma_{\mu \theta_2 \phi_2 \psi_2}\eta= -\bigg(\cos\left(\frac{\mu}{2}\right)+ \sin\left(\frac{\mu}{2}\right)\hat{\gamma}\Gamma_{\mu}\bigg)\eta,
\eeq
which breaks supersymmetry by half, leaving 16 real supercharges preserved by the reduction to Type IIA.
In fact the projection also makes the Killing spinor independent of $(\theta_1,\phi_1,\psi_1)$ in the frame of eq. \eqref{eq: M-theoryFrame} and independent of $\psi_1$ in any frame in which the Hopf isometry of $S^3_1$ is manifest. These are precisely the conditions for supersymmetry to be unbroken under $SU(2)$ and $U(1)$ T-duality transformations respectively \cite{Kelekci:2014ima,Hassan:1999bv}. So 16 supercharges will remain in Type  IIB after either of these duality
transformations, enough for
this background to be dual to  a  three-dimensional $\mathcal{N}=4$ SCFT.

\subsection{Reduction of $\mathbb{Z}_k$ orbifold to IIA}
Let us now proceed with the reduction on $\psi_2$ with a slight generalisation.  Let us reduce the $\mathbb{Z}_k$ orbifold of $S_2^3$. This has the effect of generating a stack of $k$ D6 branes in Type IIA while leaving the supersymmetry arguments unchanged. 

Taking the $\mathbb{Z}_k$ orbifold, amounts to sending $S^3_2\to S^3_2/\mathbb{Z}_k$ in  eq. \eqref{eq: ads4s7} with,
\beq
4ds(S^3_2/\mathbb{Z}_k)= d\theta_2^2+ \sin^2\theta_2 d\phi_2^2+ \frac{4}{k^2}\left(d\psi_2+\frac{k}{2}\cos\theta_2 d\phi_2\right)^2,
\eeq
where $\psi_2$ now has period $2\pi$.
Setting $l_p=\alpha'=g_s=1$ leads to the type IIA solution,
\begin{align}\label{eq:IIA sol}
ds^2_{IIA}&=e^{\frac{2}{3}\phi_0}\cos\left(\frac{\mu}{2}\right)\bigg[ ds^2(AdS_4) +4L^2\bigg(\frac{1}{4}d\mu^2+  \sin^2\left(\frac{\mu}{2}\right)ds^2(S^3_1) + \frac{1}{4}\cos^2\left(\frac{\mu}{2}\right)ds^2(S^2_2)\bigg)\bigg],\nn\\[2mm]
F_4 &= \frac{3}{L} \text{Vol}(AdS_4),~~~ F_2 = -\frac{k}{2} \text{Vol}(S^2_2),~~~ e^{2\Phi_0}= e^{2\phi_0}\cos^3\left(\frac{\mu}{2}\right),~~~ e^{\frac{2}{3}\phi_0} = \frac{2L}{k}.
\end{align}
The reduction has generated a singularity at $\mu=\pi$, but this has a physical interpretation, it is due to the $k$ D6 branes mentioned earlier.  Indeed close to $\mu=\pi$ the metric has the form,
\beq
ds^2 \sim\frac{e^{2\phi_0/3}}{2}\bigg[\sqrt{\nu}\bigg(ds^2(AdS_4)+ 4L^2 ds^2(S^3_1)\bigg)+\frac{L^2}{4\sqrt{\nu}}\bigg(d\nu^2+\nu^2ds^2(S^2_2)\bigg)\bigg],~~~e^{\Phi}\sim \frac{e^{\phi_0}\nu^{3/4}}{2\sqrt{2}},
\eeq
for $\nu=(\pi-\mu)^2$. We see that the reduction has generated $D6$ branes that extend along $AdS_4$, wrap $S^3_1$ and are localised at $\mu=\pi$.
Of course this was to be expected as D6 brane singularities are always generated anywhere the M-theory circle shrinks to zero size.

Before moving on, let us quote the D-brane charges,
\beq
N_{D2}=\frac{1}{2\kappa_{10}^2T_2}\int\star F_4= \frac{2L^6}{k\pi^2},~~~Q_{D6}=\frac{1}{2\kappa_{10}^2T_4}\int_{S^2_2} F_2=k.
\eeq
In our conventions $2 \kappa_{10}^2 T_{Dp}= (2\pi)^{7-p}$. We thus set
\beq\label{eq: Lrule1}
L^6 = \frac{k \pi^2 N_{D2}}{2},
\eeq
to have integer D2 brane charge. We find the expected number of D6 branes.

\subsection{The non-Abelian T-dual solution}
We now present the solution that will be the main focus of this work, which is the result of performing a non-Abelian T-dual transformation on the $S^3_1$ of  eq. \eqref{eq:IIA sol}. Using the rules in \cite{Kelekci:2014ima}, and parametrising the T-dual coordinates in terms of spherical  coordinates $(r,S^2_1)$, we generate the NS sector,
\begin{align}\label{eq: IIBmetric}
ds^2_{IIB} =& e^{\frac{2}{3}\phi_0}\cos\left(\frac{\mu}{2}\right)\bigg[ ds^2(AdS_4) +L^2\bigg( d\mu^2+ \frac{k^2}{L^6 \sin^2\left(\mu\right)}dr^2+  \cos^2\left(\frac{\mu}{2}\right)ds^2(S^2_2)\bigg)\bigg]\nn\\
+&\frac{L^6}{k^2 \Delta}r^2 \sin^2\left(\frac{\mu}{2}\right)\sin^2\left(\mu\right)ds^2(S^2_1),\nn\\[2mm]
B_2=& \frac{L^3 }{k\Delta}r^3\sin\left(\frac{\mu}{2}\right)\sin\left(\mu\right)\text{Vol}(S_1^2),~~~  e^{2\Phi}=\frac{1}{\Delta} e^{2\phi_0}\cos^3\left(\frac{\mu}{2}\right),
\end{align}
where we have introduced
\beq
\Delta = \frac{L^3}{k^3}\sin\left(\frac{\mu}{2}\right)\sin\left(\mu\right)\bigg(k^2r^2 + L^6\sin^2\left(\frac{\mu}{2}\right)\sin^2\left(\mu\right) \bigg).
\eeq
The solution is completed with the RR fluxes,
\begin{align}\label{eq:NATD_Fluxes}
F_3=& \frac{1}{4}\text{Vol}(S^2_2)\wedge d\bigg(k r^2-\frac{L^6}{k}\big(\cos^2(\mu)-3)\cos(\mu)\big)\bigg),\nn\\[2mm]
F_5=& \text{Vol}(AdS_4)\wedge d\bigg(\frac{L^5}{4 k^2}\big(\cos(2\mu)-4\cos(\mu)\big)-\frac{3}{2 L}r^2\bigg)\nn\\
-&\frac{L^9}{4k^2\Delta}r^2\sin^3(\mu)\sin\left(\frac{\mu}{2}\right) \text{Vol}(S^2_1)\wedge\text{Vol}(S^2_2)\wedge \bigg(3 r\sin(\mu)d\mu+2 \sin^2\left(\frac{\mu}{2}\right)dr\bigg).
\end{align}
We have explicitly checked that the background in eqs.(\ref{eq: IIBmetric})-(\ref{eq:NATD_Fluxes}) solves the Type IIB Supergravity equations of motion, which is also implied by the result of the paper \cite{Itsios:2012dc}.

As is common to all backgrounds generated through an $SU(2)$ non-Abelian T-duality transformation,  
this solution incorporates a non-compact $r$-direction. Moreover, this solution has two singularities. The first lies at $\mu=\pi$ and is inherited from  the stack of D6 branes in IIA. Indeed, close to $\mu=\pi$ one finds
\beq\label{eq:NATD_D5_smeared}
ds^2 \sim\frac{e^{2\phi_0/3}}{2}\bigg[\sqrt{\nu}\bigg(ds^2(AdS_4) + L^2ds^2(S^2_1)\bigg) +\frac{ L^2 }{4\sqrt{\nu}}\bigg(d\nu^2+ d\tilde{r}^2+\nu^2 ds^2(S^2_2)\bigg)\bigg],~~~~~~ e^{\Phi}\sim \frac{2\sqrt{\nu}}{L^3 \tilde{r}}
\eeq
where we have defined $\tilde{r}=2k/L^2 r$  and $\nu=(\pi-\mu)^2$. This is almost the behaviour of the smeared D5 stack one would generate under Hopf fibre T-duality along $\psi_1$.   The $r$-dependence of the dilaton however modifies this. Recalling that the dilaton is determined by a one loop effect in T-duality, which essentially amounts to imposing that $e^{-2\Phi}\text{Vol}(\mathcal{M}_I)$ (where $\mathcal{M}_I$ is the submanifold where the duality is performed) is duality invariant, the $r$ factor has its origin in the different volumes of the original and non-Abelian T-dual submanifolds, which are respectively $S^3$ and $\mathbb{R}^3$. This is manifest when we parametrise the volume of $\mathbb{R}^3$ in spherical coordinates $(r, S_1^2)$, where $r$ is the radial direction.
The second singularity at $\mu=0$ is also unsurprising, since we have dualised on a sphere whose radius vanishes at this point. We indeed obtain the non-Abelian T-dual analogue of smeared NS5 branes, since close to $\mu\sim 0$ we have,
\beq\label{eq:NATD_NS5_smeared}
ds^2\sim e^{2\phi_0/3}\bigg[ds^2(AdS_4)+  L^2 ds^2(S^2_2)+ \frac{L^2}{4\nu}\bigg(d\nu^2+ d\tilde{r}^2+\nu^2 ds^2(S^2_1)\bigg)\bigg],~~~e^{\Phi}\sim\frac{8}{L^3\sqrt{\nu}\tilde{r}} 
\eeq
where now we have defined $\nu=\mu^2$ and once more it is the dependence of the dilaton on $r$ that makes this deviate from the conventional $(\sqrt{\nu})^{-1}$ behaviour. 

\begin{figure}
\centering
\includegraphics[scale=0.6]{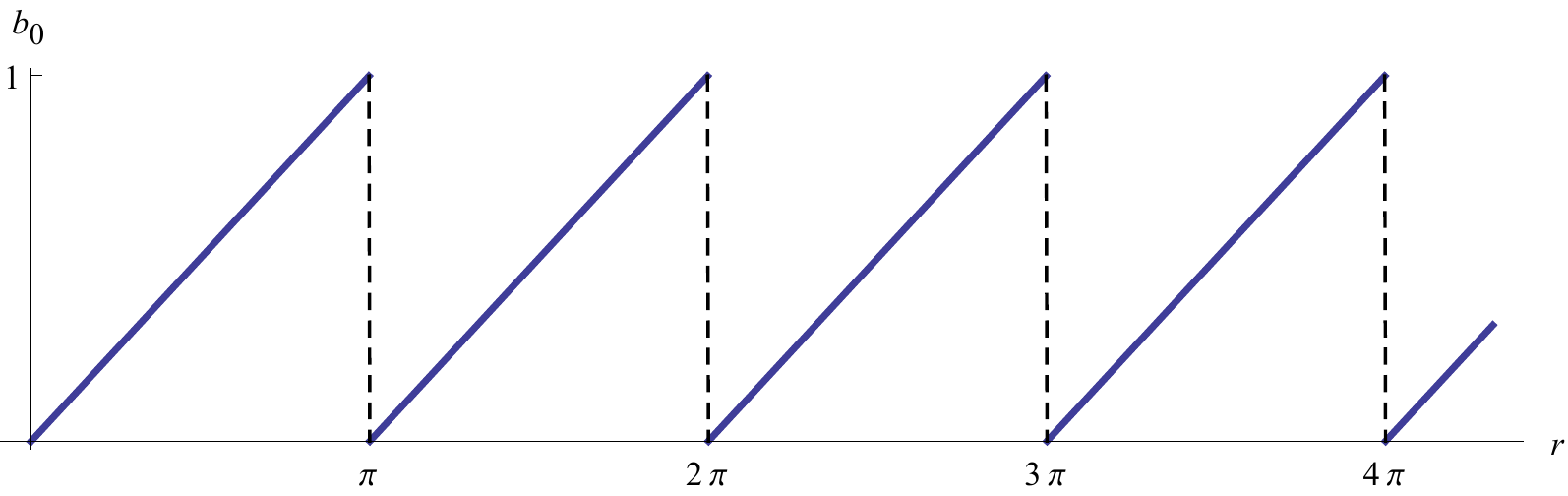}
\caption{$b_0$ as a function of $r$}
\label{b0}
\end{figure}

As previously discussed in other non-Abelian T-dual examples---see \cite{Lozano:2013oma,Lozano:2014ata,Macpherson:2014eza,Macpherson:2015tka},
the behaviour of the solution close to the location of the NS5-branes brings in interesting information.
Close to $\mu=0$  we have $B_2= r \text{Vol}(S^2_1)$, with the metric spanned by $(\mu,S^2_1)$ becoming a singular cone, which defines a 2-cycle. This means that we must ensure that on this cycle $S^2_1$, the quantity
\beq\label{eq:b0NATD}
b_0=\frac{1}{4\pi^2}\int_{S^2_1} B_2
\eeq
satisfies $b_0\in [0,1]$ over the infinite range of $r$. This is achieved
by performing a large gauge transformation $B_2 \to B_2-n \pi\text{Vol}(S^2_1)$ every time we cross $r=n\pi$ for $n=0,1,2,...$ so that $b_0$ is a piecewise linear periodic function as illustrated in Figure \ref{b0}. 
In this way $r$ is naturally partitioned into intervals of length $\pi$, with different brane content in each one of them, as the study of the Page charges reveals. 
Indeed, there are two charges defined on compact sub manifolds, 
\beq\label{eq:NATD_charges}
N_{D5}=\frac{1}{2\kappa_{10}^2T_5}\int_{\Sigma_1}F_3 = \frac{L^6}{k\pi},~~~N_{D3}=\frac{1}{2\kappa_{10}^2T_3}\int_{\Sigma_2}\big(F_5-B_2\wedge F_3\big) = n N_{D5},
\eeq
where $\Sigma_1 = (\mu,S^2_2)$, $\Sigma_2=(\mu,S^1_2,S^2_2)$. Thus, we need to tune
\beq
L^6= k N_{D5}\pi .
\label{rescaledspace}
\eeq
Notice that $N_{D3}$ is not globally defined. Instead its value depends on which interval we consider.  
In addition to this there are three charges that are defined on the non compact sub-manifolds,
\beq
 \tilde{\Sigma}_1=(r,S^2_1),~~~\tilde{\Sigma}_2 = (r,S^2_2),~~~ \tilde{\Sigma}_3=(r,S^2_1,S^2_2).
\eeq
We take the non compact $r$ to be indicative of an infinite linear quiver, as shown for a related $AdS_5$ example in \cite{Lozano:2016kum}. 
We calculate the charges in the interval $r\in[n\pi,(n+1)\pi]$ and  find,
\begin{align}\label{eq:NATD_charges2}
N_{NS5}&=\frac{1}{2\kappa_{10}^2T_{NS5}}\int_{S^2_1}\int_{n\pi}^{(n+1)\pi} dr\, H_3=1,\nn\\[2mm]
k_{D5}&=-\frac{1}{2\kappa_{10}^2T_5}\int_{S^2_2}\int_{n\pi}^{(n+1)\pi} dr \, F_3 =(1+2n) \frac{ k \pi}{4}\equiv (2n+1) k_0.
\end{align}
Notice that the parameter $k$, originally quantised in the Type
IIA solution needs to be re-quantised according to $k\pi=4 k_0$,
after the non-Abelian T-duality. The same happens to the size of the space $L$ as shown in eq.(\ref{rescaledspace}).

We can also compute
\bea
k_{D3}&=-\frac{1}{2\kappa_{10}^2T_3}\int_{S^2_1\times S^2_2}\int_{n\pi}^{(n+1)\pi} dr\,\big(F_5-B_2\wedge F_3\big) = (3n+2) \frac{k_0}{3},\nn
\eea
but this last one will not be relevant in our analysis below.
Notice that all these charges are integer provided $\frac{k\pi}{12}=\frac{1}{3}k_0$ is an integer.

\begin{figure}
\centering
\includegraphics[scale=0.45]{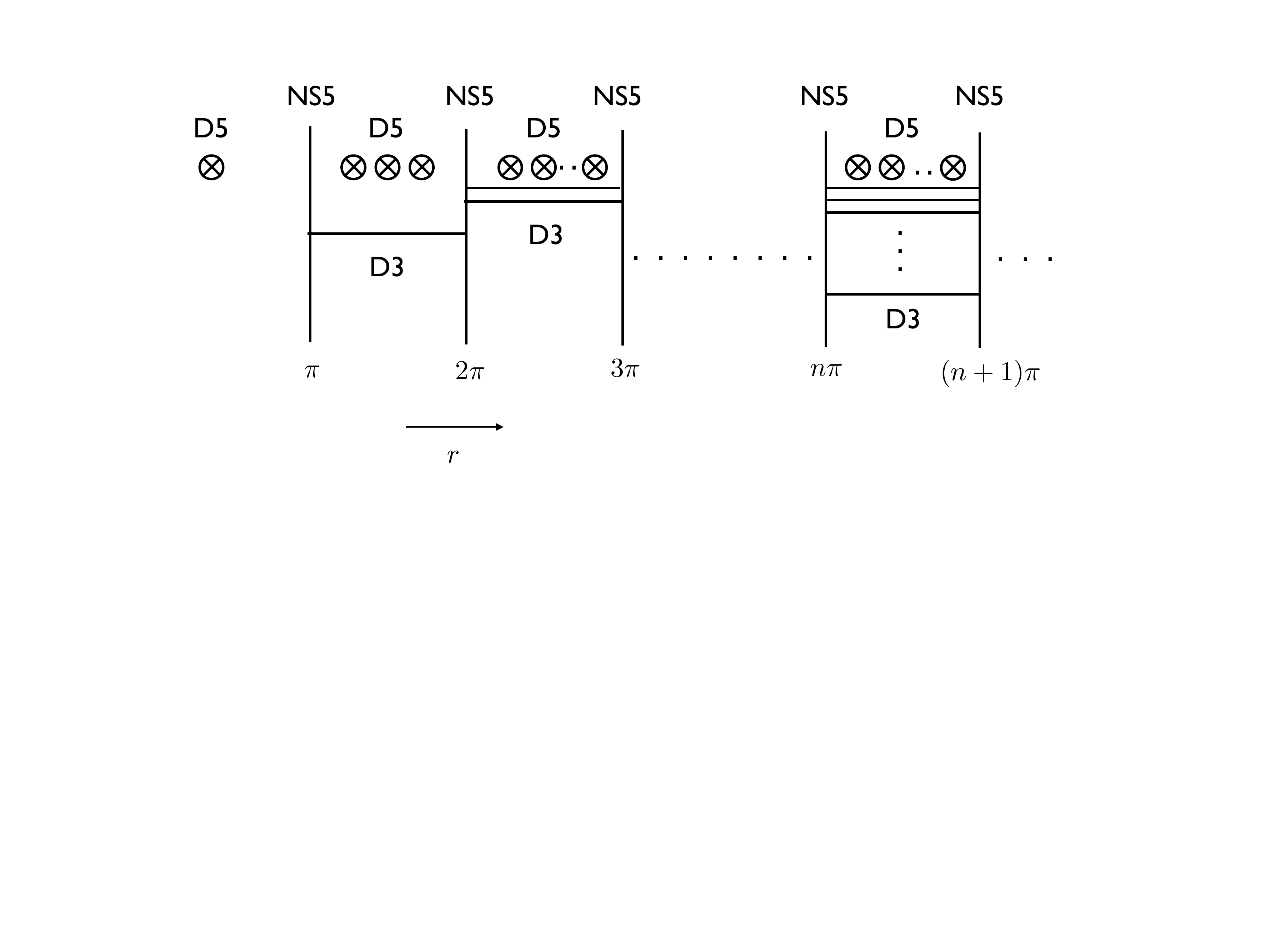}
\vspace{-6.5cm}
\caption{(NS5, D3, D5) brane set-up. The number of D3-branes is given in $N_{D5}$ units and that of D5-branes in $k_0$ units.}
\label{stacks1}
\end{figure}

The previous analysis suggests a (NS5, D3, D5) brane set-up in which NS5-branes wrapped on $AdS_4\times S^2_2$ are located at $\mu=0$, $r=\pi, 2\pi,\dots,n\pi$, with $n$ running to infinity, and there are $n N_{D5}$ D3-branes, extended on $(\mathbb{R}^{1,2}, r)$ stretched among the $n$'th and $(n+1)$'th NS5's. On top of this, $(2n+1) k_0$ D5-branes, wrapped on $AdS_4\times S^2_1$ and located at $\mu=\pi$, lie between the $n$'th and $(n+1)$'th NS5-branes. This brane set-up is depicted in Figure \ref{stacks1}. After we recall some basic properties of 3d $\mathcal{N}=4$ CFTs and their holographic duals, following  \cite{Gaiotto:2008ak,Assel:2011xz}, we will make a concrete proposal for the field theory living on this brane configuration.

\section{Aspects of 3d ${\cal N}=4$ CFTs and their holographic duals}\label{seccion3}

In this section we recall the basic aspects of the three dimensional $\mathcal{N}=4$ field theories studied in \cite{Gaiotto:2008ak} and of their holographic duals, derived in \cite{Assel:2011xz,Aharony:2011yc}. We start with the field theory description.

\subsection{3d ${\cal N}=4$ CFTs} 

The study of the moduli space of ${\cal N}=4$ SYM in four dimensions defined on an interval
 with SUSY preserving boundary conditions,  lead
 Gaiotto and Witten \cite{Gaiotto:2008ak} to introduce a family of 3d quantum field theories ---named $T_{\rho}^{\hat{\rho}} (SU(N))$, characterised by an integer $N$ and two partitions of it, denoted $\rho$ and $\hat{\rho}$.
 From $(N$, $\rho$, ${\hat \rho})$ it is possible to read the data defining the UV of these theories, namely, the gauge group $G=U(N_1)\times....\times U(N_k)$, the bi-fundamental fields transforming in the
 $(N_i,\bar{N}_{i+1})$ representations, and the fundamental matter, transforming under $U(M_i)$  for each gauge group. 
 
Given a list of positive numbers [$l_1\geq l_2\geq....\geq l_p$], one can define a partition $\rho$ of $N$ by $N=\sum_{r=1}^p M_r l_r$. The numbers $M_r$, which indicate how many times the different integers $l_r$ appear in the partition, give the ranks of the fundamental matter groups in the field theory. Similarly, one can define a second partition $\hat{\rho}$, consisting of the numbers
$[\hat{l}_1\geq {\hat l}_2 \geq....\geq \hat{l}_{\hat{p}}]$, with multiplicities $\hat{M}_r$, such that
$N=\sum_{r=1}^{\hat{p}} \hat{M}_r \hat{l}_r$. From these partitions the ranks of the different $U(N_i)$  gauge groups are computed from the expressions,
\begin{equation}
N_i=\sum_{s=1}^{i} (m_s-{\hat l}_s)\, ,
\end{equation}
where $m_s$ denotes the number of terms that are equal or bigger than a given integer $s$ in the decomposition $N=\sum_{r=1}^p M_r l_r$. 

 Gaiotto and Witten \cite{Gaiotto:2008ak} conjectured that the condition for these three-dimensional field theories to flow to a conformal fixed point is (schematically)
  $\rho^T\geq\hat{\rho}$. More specifically, this condition means that 
  \begin{equation}
  \sum_{s=1}^{i} m_s \geq \sum_{s=1}^i {\hat l}_s \quad \forall i=1,\dots {\hat p}\, .
  \end{equation}
  Associating a Young Tableau with rows of lengths $[l_1, \dots, l_p]$ to the partition $\rho$ and one with columns of lengths  $[\hat{l}_1,....,\hat{l}_{\hat{p}}]$ to the partition ${\hat \rho}$, this condition means that the number of boxes in the first i-rows of the Young Tableau associated to $\rho^T$ must be larger or equal than 
 the corresponding number in the Tableau associated to $\hat{\rho}$.
 In those cases in which the equality holds, that is,
 \begin{equation}
  \sum_{s=1}^{i} m_s = \sum_{s=1}^i {\hat l}_s \quad {\rm for \,\,\, some}\,\, i\, ,
  \end{equation}
some gauge groups have zero rank, and the quiver becomes disconnected.

 The quantum theory defined by $T_{\rho}^{\hat{\rho}} (SU(N))$ has Coulomb and Higgs branches of vacua, while the  theory defined by $T_{\hat{\rho}}^{\rho} (SU(N))$ 
has the same moduli space, but with the Coulomb and Higgs vacua interchanged. Both theories are conjectured to flow to the same IR fixed point, which is a reflection of mirror symmetry. 
The three-dimensional  CFT that appears at low energies is invariant under $SO(2,3)$--reflecting the conformality in 3d, and $SO(4)\sim SU(2)_L\times SU(2)_R$--reflecting the R-symmetry of $\mathcal{N}=4$ SUSY in 3d. This field theory can be nicely realised through a Hanany-Witten  \cite{Hanany:1996ie} set-up consisting of $p$ D5 branes and $\hat{p}$ NS5 branes with D3 branes stretched between them. This brane set-up is shown in Table \ref{brane_set-up}.

\begin{table}[h]
\label{brane_set-up}
\begin{center}
\begin{tabular}{| l | c | c | c | c| c | c| c | c| c | c |}
 \hline
    & 0 & 1 & 2 & 3 & 4 & 5 & 6 & 7 & 8 & 9 \\ \hline
 D3 & x & x & x & x &   &   &   &   &   &   \\ \hline
 D5 & x & x & x &   & x & x & x &   &   &   \\ \hline
NS55 & x & x & x &   &   &   &   & x & x & x \\ \hline
\end{tabular} 
\end{center}
\caption{Hanany-Witten brane set-up corresponding to the $\mathcal{N}=4$ 3d theory.}   
\end{table}   
%

\noindent The $x_3$-direction on which D3 branes stretch is of finite size, thus giving rise at long distances to a three-dimensional QFT  on $[0,1,2]$. The $SU(2)_L\times SU(2)_R$ R-symmetry is associated with rotations in the $[4,5,6]$ and $[7,8,9]$ directions. In turn, the $l_1\geq l_2\geq....\geq l_p$ and $\hat{l}_1\geq {\hat l}_2 \geq....\geq \hat{l}_{\hat{p}}$ numbers that define the partitions $(\rho, {\hat \rho})$ are respectively, the linking numbers associated to the $p$ D5 and ${\hat p}$ NS5 branes. These are defined by
\bea
& & l_{D5,a}=l_a = - n_a + R_a^{NS5}, \;\;\;\; a=0,....,p, \nonumber\\
& & \hat{l}_{NS5,b}=\hat{l}_b= n_b + L_b^{D5}\;\;\;\; b=1,....,\hat{p},\nonumber
\eea
where $n_a$ is the net number of D3  branes ending on the given five brane (number of D3 on the right - number of D3 on the left).
In turn, $R_a^{NS5}$ is the number of NS5 branes to the {\it right} of a given D5 brane, while 
 $L_b^{D5}$ is the number of D5 branes to the {\it left} of a given NS5 brane. The multiplicities of each linking number, $M_r$, ${\hat M}_r$, are thus the number of branes in the corresponding stack of D5 or NS5 branes. 
 
 \subsection{The ABEG dual geometries}\label{seccion4}
 
 Following the formulation initiated in  \cite{D'Hoker:2007xz,D'Hoker:2007xy}, the authors of  \cite{Assel:2011xz,Aharony:2011yc} proposed  that the supergravity solutions associated to the three dimensional ${\cal N}=4$ CFTs that we just described are fibrations of $AdS_4\times S^2\times S^2$ over a Riemann surface $\Sigma_2$. We will refer to these geometries as ABEG geometries for brevity.
These solutions have manifest $SO(2,3)\times SU(2)_L\times SU(2)_R$ symmetry, and can be completely determined from two harmonic functions $h_1(z,\bar{z}), h_2(z,\bar{z})$, defined on the Riemann surface $\Sigma_2$.
From  the functions $h_1$, $h_2$, the background and fluxes are given by,
\begin{align}\label{eq: Ansatz}
ds^2&=\lambda^2 ds^2(AdS_4)+ \lambda_1^2 ds^2(S_1^2)+\lambda_2^2 ds^2(S_2^2)+ ds^2(\Sigma_2),\nn\\[2mm]
H_3 &= d(b_1)\wedge \text{Vol}(S^2_1),~~~~F_3 = d(b_2)\wedge\text{Vol}(S^2_2),~~~ ds^2(\Sigma_2) = 4\tilde{\rho}^2 |dz|^2\nn\\[2mm]
F_5&={4}(1+\star) f\wedge \text{Vol}(S^2_1)\wedge \text{Vol}(S^2_2),
\end{align}
where
$\tilde{\rho}^2,\lambda,\lambda_1,\lambda_2,b_1,b_2$ and the dilaton $e^{\Phi}$  are real functions and $f$ denotes a 1-form on $\Sigma_2$, explicitly written below.
These functions can be written in a compact form from $h_1$, $h_2$ using,
\begin{align}\label{eq:WN1N2}
W&= \partial_z h_1\partial_{\bar{z}} h_2+\partial_{\bar{z}} h_1\partial_z h_2,~~~ X= i(\partial_z h_1\partial_{\bar{z}} h_2-\partial_{\bar{z}} h_1\partial_z h_2)\nn\\[2mm]
N_1&=2 h_1 h_2 |\partial_z h_1|^2- h_1^2 W,~~~~~
N_2=2 h_1 h_2 |\partial_z h_2|^2- h_2^2 W,\end{align}
as,
\begin{align}\label{eq:ansatzrules}
&\tilde{\rho}^2= \frac{\sqrt{N_2 |W|}}{h_1 h_2},~~~\lambda^2= 2 \sqrt{\frac{N_2}{|W|}},~~~\lambda^2_1=2e^{\Phi}h_1^2\sqrt{\frac{|W|}{N_1}},~~~\lambda^2_2=2h_2^2\sqrt{\frac{|W|}{N_2}},\nn\\[2mm]
&b_1=2 h^D_2+2 h_1^2 h_2 \frac{X}{N_1},~~~b_2=-2 h^D_1+2 h_1 h_2^2 \frac{X}{N_2},\;\;\;e^{2\Phi}=\frac{N_2}{N_1}.
\end{align}
Here $h^D_1,h^D_2$ are the harmonic duals of $h_1,h_2$, defined such that $h^D_1+i h_1$ and $h_2-i h^D_2$ are holomorphic functions.
{Notice that we are working in string frame, hence some factors of the dilaton  differ from \cite{D'Hoker:2007xy,Aharony:2011yc}, which use Einstein frame.}
Finally,  the 1-form $f$ is given by,
\beq
\label{efe}
f = 2 \text{Im}\bigg(\bigg[3i \bigg(h_1 \partial_z h_2- h_2\partial_z h_1\bigg)+ \partial_z\left(h_1 h_2 \frac{X}{W}\right)\bigg]\frac{\lambda_1^2\lambda_2^2}{\lambda^4}dz\bigg).
\eeq 

\begin{figure}
\centering
\includegraphics[scale=0.7]{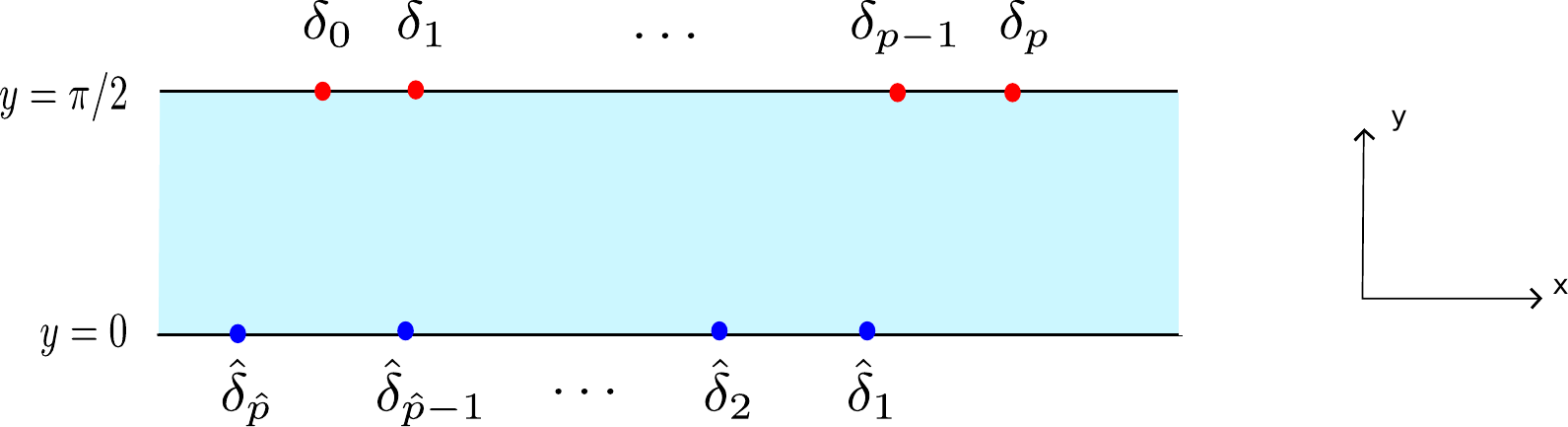}
\caption{5-branes distribution along the strip, parameterised by $z=x + iy$.}
\label{deltas_ref}
\end{figure}

It was shown in 
 \cite{Assel:2011xz} that the two harmonic functions $h_1$, $h_2$ that encode the supergravity solution as shown above, can be determined from the (D5, NS5, D3) brane set-up associated to the  
$T_{\rho}^{\hat{\rho}} (SU(N))$ theory. Defining the sets of numbers  $[N_5^a,\delta_a]$ and 
 $[\hat{N}_5^{b},\hat{\delta}_b]$, denoting respectively the number of branes at each stack and the position of this stack, for D5 and NS5 branes, and taking $\Sigma_2$ as the strip defined by 
$-\infty<Re[z]<\infty$ and $0\leq Im[z]\leq \frac{\pi}{2}$ \footnote{This choice of strip is consistent for linear quivers (see \cite{Assel:2011xz,Assel:2012cj}).}, the $h_1$, $h_2$ functions are given by,
\bea
 h_1= -\frac{1}{4}\sum_{a=1}^{p} N_5^{a}\log\tanh (\frac{i\frac{\pi}{2} +\delta_a-z}{2}) + cc,\;\;\; 
 h_2= -\frac{1}{4}\sum_{b=1}^{\hat{p}} \hat{N}_5^{b} \log\tanh (\frac{z-\hat{\delta}_b}{2})+cc.
\label{h's} 
 \eea
These expressions exhibit logarithmic singularities at the locations of the stacks of D5-branes, at $z=\delta_a+i\pi/2$, for $h_1$, 
and at the locations of the NS5-branes  $z={\hat \delta}_b$, for $h_2$. 
The brane distribution is depicted in Figure \ref{deltas_ref}. The Laplace problem that these functions solve must be complemented by conditions on the boundaries of the Riemann surface \cite{D'Hoker:2007xz,D'Hoker:2007xy},
 \bea
 h_1\big\lvert_{\text{Im}[z]=0}= \partial_\perp h_2\big\lvert_{\text{Im}[z]=0}=0,\;\;\;\;  h_2\big\lvert_{\text{Im}[z]=\frac{\pi}{2}}= \partial_\perp h_1\big\lvert_{\text{Im}[z]=\frac{\pi}{2}}=0,
 \label{bcs}
 \eea
 where  $\partial_\perp=\partial_z-\partial_{\bar z}$, which the $h_1$ and $h_2$ in eq.(\ref{h's}) satisfy.  

From the expressions for $h_1$, $h_2$ in eq.(\ref{h's}), the fluxes in eq.(\ref{eq: Ansatz}) can be calculated using eqs.(\ref{eq:WN1N2}) and (\ref{eq:ansatzrules}). 
The associated charges are defined as,
 \beq
 N_5^a=\frac{1}{2\kappa_{10}^2 T_{D5} }\int_{I\times S_2^2}F_3,\;\;\;\; \hat{N}_5^b=\frac{1}{2\kappa_{10}^2 T_{NS5}}\int_{\hat{I}\times S_1^2}H_3,\label{chargesnsd5}
 \eeq
 where the 3-cycles, defined in   \cite{Assel:2011xz}, consist of a shrinking sphere times an interval $I$ or $\hat{I}$, that semi-circles
the position of the singularity at $\delta_a$ or $\hat{\delta}_b$. As we discussed, $(N_5^a, \hat{N}_5^b)$ should be identified with the multiplicities $(M_r,\hat{M}_r)$ in the two partitions $\rho, {\hat \rho}$.
 
Similarly, it is possible to define two Page charges associated to D3 branes, one being the S-dual of the other:
 \beq
  N_{3}^a= \frac{1}{2\kappa_{10}^2 T_{D3} }\int_{I \times S_1^2\times S_2^2}[F_5- B_2\wedge F_3] ,\;\;\;\;  
  \hat{N}_{3}^b= \frac{1}{2\kappa_{10}^2 T_{D3}}\int_{\hat{I}\times S_1^2\times S_2^2}[F_5+C_2\wedge H_3].
  \label{pagesd3}
 \eeq
  These charges are well defined whenever the potential $B_2$ or $C_2$ entering in their expression is well-defined, that is, away from the positions where  the NS5 or D5 branes are located. From these and the previous charges, the linking numbers associated to the D5 and NS5 branes can be determined as  \cite{Assel:2011xz},
  \bea
 l_a=-\frac{N_{3}^a}{N_5^a}= \frac{2}{\pi}\sum_{b=1}^{\hat{p}} \hat{N}_5^b \arctan( e^{\hat{\delta}_b-\delta_a}),\;\;\;\;
 \hat{l}_b=\frac{\hat{N}_{3}^b}{\hat{N}_5^b}= \frac{2}{\pi}\sum_{a=1}^{{p}} {N}_{5}^a \arctan( e^{\hat{\delta}_b-\delta_a}).
\label{linkings}
\eea
 As expected, they satisfy
 \beq\label{eq:defN}
 N=\sum_{b=1}^{\hat{p}} \hat{N}_5^b \,\hat{l}_b=\sum_{a=1}^p N_5^a \,l_a. 
 \eeq

Finally, in \cite{Assel:2012cp} a special limit of the general expressions for $h_1$ and $h_2$ given in eq.(\ref{h's})  was considered.
In this limit, the NS5-branes and D5-branes located at the two boundaries of the strip, ${\rm Im}~ z=0$, ${\rm Im} ~z=\pi/2$, are positioned at infinite values of ${\rm Re}~ z$. This limit will be useful when we discuss the realisation of the non-Abelian T-dual solution as an ABEG geometry. Specifically, it was shown in  \cite{Assel:2012cp} that
if $\delta_a\rightarrow \infty$ and ${\hat \delta}_b\rightarrow -\infty$  one can approximate eq. \eqref{h's} by,
\begin{eqnarray}
\label{h1h2}
h_1&=& \sin{y}\sum_{a=1}^p N_5^{a} e^{x-\delta_a}+\dots\quad {\rm if} \,\,x<\delta_1,
\nonumber\\
&=& \sin{y}\sum_{a=i}^p N_5^{a} e^{x-\delta_a}+\dots\quad {\rm if} \,\,\delta_{i-1}<x<\delta_{i},\nonumber\\
h_2&=& \cos{y}\sum_{\hat{b}=1}^{\hat p} \hat{N}_5^{b} e^{\hat{\delta}_b-x}+\dots \quad {\rm if} \,\,x>{\hat \delta}_1, \nonumber\\
&=& \cos{y}\sum_{\hat{b}=i}^{\hat p} \hat{N}_5^{b} e^{\hat{\delta}_b-x}+\dots \quad {\rm if} \,\,{\hat \delta}_{i-1}>x>{\hat \delta}_i ,
\end{eqnarray}
where the strip is parameterised by $z=x+iy$.
Notice that these expressions still satisfy the boundary conditions in eq.(\ref{bcs}).

\section{The Type IIB $\mathcal{N}=4$ $AdS_4$ solution and CFT}\label{seccion5}

After we have discussed the basic ingredients of 3d $\mathcal{N}=4$ CFTs and their duals, we can go back to our brane configuration, discussed at the end of Section \ref{sec:sugra_solutions}, and make a concrete proposal for the CFT associated to the brane set-up depicted in Figure \ref{stacks1}. 

Restricting the $r$ direction to lie between zero and $r=(n+1)\pi$,  we have a total number of $n+1$ NS5-branes (see Figure \ref{stacks2}). In order to have a field theory that flows to a non-trivial infrared fixed point (see below) we need to add $(n+1)N_{D5}$ D3-branes ending on the $(n+1)$'th NS5-brane from the right. This is achieved inserting a stack of $(n+1)N_{D5}$ D5-branes to the right of the $(n+1)$'th NS5-brane, each one connected to this NS5-brane by a D3-brane. In turn, this is equivalent up to a Hanany-Witten move \cite{Hanany:1996ie} to just taking the $n$'th stack of D5-branes with $(2n+1) k_0+(n+1)N_{D5}$ branes.
This  field theoretical completion of the quiver has the geometric counterpart of making finite the range of the $r$-coordinate.

Thus, in the notation of ABEG, we have $p=n+1$ and the multiplicity of D5 branes is,
\begin{equation}
\label{N5a}
N_5^{a}=(2a+1)k_0\, , a=0,\dots, n-1\, \qquad N_5^{n}=(2n+1)k_0+(n+1) N_{D5}\, .
\end{equation}

We can now compute the linking numbers associated to the five branes in the Hanany-Witten set-up. 
These provide an invariant way of encoding the brane configuration, since they do not change under Hanany-Witten moves. 
The linking numbers associated to the D5-branes are given by ,
\begin{equation}
l_a=-n_a+R_a^{NS5},
\end{equation}
where $n_a$ denotes the net number of D3-branes ending on the $a$'th stack of D5-branes and $R_a^{NS5}$ the number of NS5-branes located at its right. 
\begin{figure}
\centering
\includegraphics[scale=0.45]{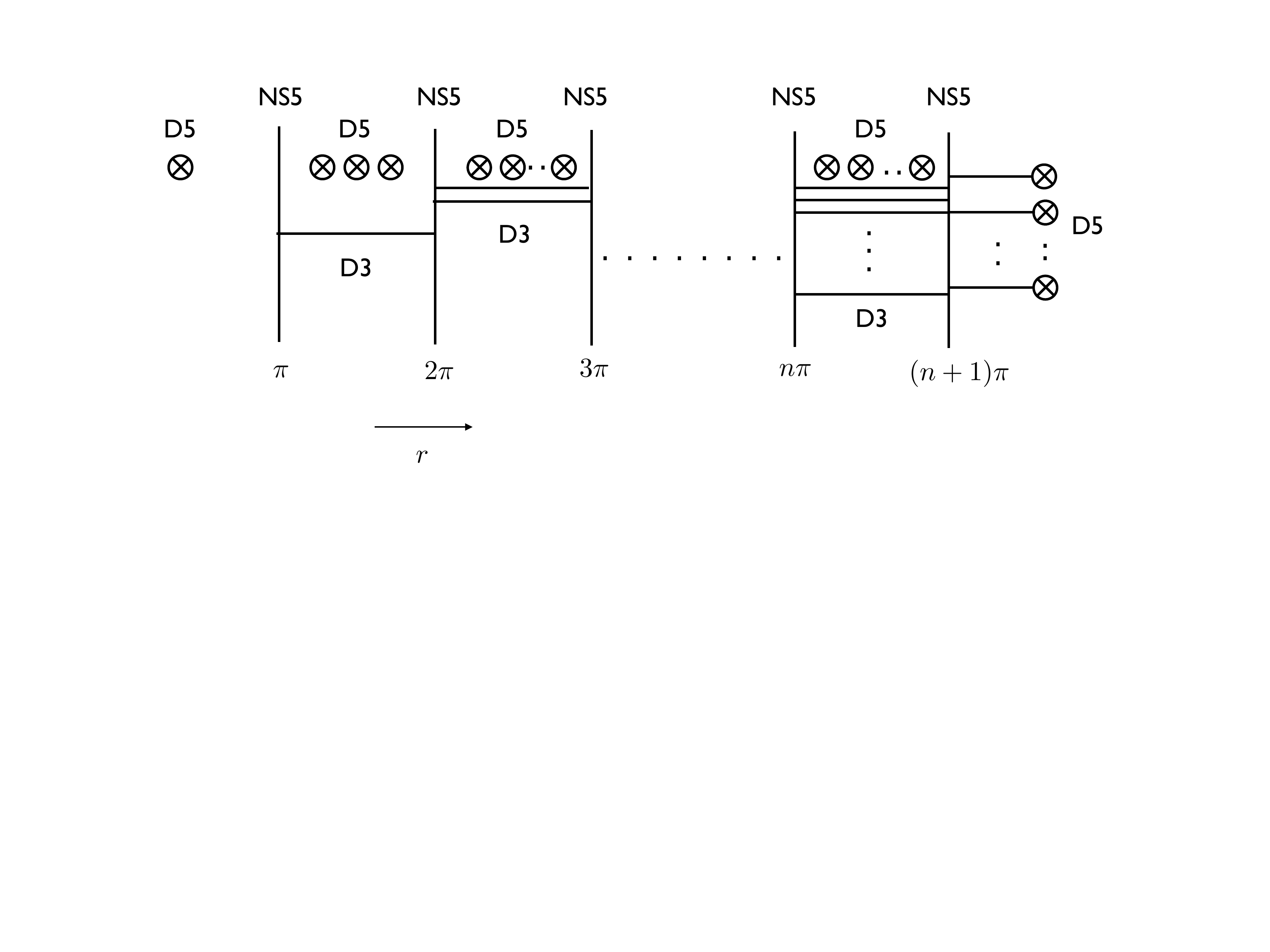}
\vspace{-6.5cm}
\caption{Completed (NS5, D3, D5) brane set-up.}
\label{stacks2}
\end{figure} 
For our brane set-up we find, 
\begin{equation}
\label{linkings1}
l_a=n+1-a\quad {\rm for}\quad a=0,1,\dots,n  .
\end{equation}
From here the total number of D3-branes
 $N$, reads
\begin{eqnarray}
\label{N}
N&=&\sum_{a=0}^{n}l_a N_5^{a}=\sum_{a=0}^n (n+1-a)(2a+1)k_0+(n+1)N_{D5}=\nonumber\\
&=&\frac{k_0}{6}(n+1)(n+2)(2n+3)+N_{D5}(n+1) .
\end{eqnarray}
Alternatively, we can compute $N$ using the NS5-branes stacks. In this case the linking numbers are computed from,
\begin{equation}
{\hat l}_b=n_b+L_b^{D5},
\end{equation}
where $n_b$ denotes once more the net number of D3-branes ending on the $b$'th NS5-brane, and $L_b^{D5}$ denotes the number of D5-branes to the left of the $b$'th NS5-brane. We find that,
\begin{equation}
\label{linkings2}
{\hat l}_b=N_{D5}+k_0 b^2  , b=1,\dots, n+1.
\end{equation}
Once can easily check that 
\begin{equation}
N=\sum_{b=1}^{n+1}{\hat l}_b N_5^{b}=\sum_{b=1}^{n+1}{\hat l}_b=\frac{k_0}{6}(n+1)(n+2)(2n+3)+N_{D5}(n+1)\, ,
\end{equation}
as in eq.(\ref{N}).
Thus, the $\rho, {\hat \rho}$ partitions associated to the brane configuration in Fig.\ref{stacks2} read 
\begin{equation}
\label{rho}
\rho: 	\qquad N=\underbrace{1+\ldots+1}_{(2n+1) k_0+(n+1)N_{D5}}+\underbrace{2+\ldots+2}_{(2n-1)k_0}\, +\ldots + \underbrace{(n+1)+\ldots +(n+1)}_{k_0},
\end{equation}
and
\begin{equation}
\label{rhohat}
{\hat \rho}: \qquad N=\underbrace{N_{D5}+(n+1)^2 k_0}_{1}+\underbrace{N_{D5}+n^2 k_0}_{1}+\ldots +\underbrace{N_{D5}+k_0}_{1}.
\end{equation}
These two partitions define the $T_\rho^{\hat{\rho}}(SU(N))$ field theory associated to our brane set-up.
Following now the work of  ABEG \cite{Assel:2011xz} we can read from eq.(\ref{rho}) the number of terms, $m_l$, that are equal or bigger than a given integer $l$,
\begin{equation}
m_1=(n+1) N_{D5}+(n+1)^2 k_0\, , \quad m_2= n^2 k_0\, , \quad \dots \quad , m_n=4 k_0,\quad m_{n+1}=k_0.
\end{equation}
From these, the condition to have a field theory that flows to a non-trivial infrared fixed point, as was conjectured in \cite{Gaiotto:2008ak}, is,
\begin{equation}
\rho^T \geq {\hat \rho} \qquad \Longleftrightarrow \qquad \sum_{s=1}^{i} m_s\,\geq \,\sum_{s=1}^{i}{\hat l}_s \qquad  \forall i=1,\ldots, n+1,
\end{equation}
where for this to hold the ${\hat l}_i$ must be ordered such that ${\hat l}_1\geq {\hat l}_2\dots \geq {\hat l}_i$. We will use 
this notation in the  rest of this section. In the present case we have,
\begin{equation}
\sum_{s=1}^{i} m_s=(n+1) N_{D5}+\sum_{q={n-i+2}}^{n+1} q^2 k_0 \, ,
\end{equation}
which is strictly larger than 
\begin{equation}
\sum_{s=1}^{i} {\hat l}_s=i N_{D5}+\sum_{q={n-i+2}}^{n+1} q^2 k_0 ,
\end{equation}
for $i=1,\ldots, n$, while 
\begin{equation}
\sum_{s=1}^{n+1} m_s=\sum_{s=1}^{n+1} {\hat l}_s .
\end{equation}
The last condition is consistent with the fact that there are $k_0$ D5-branes in the $[0,\pi]$ interval that are disconnected from the rest of the branes, thus leading to a quiver that breaks into two disconnected components.
We can also check that, consistently with our brane set-up in Figure \ref{stacks2}, the ranks of the gauge groups are given by
\begin{equation}
N_i=\sum_{s=1}^{i} (m_s-{\hat l}_s)=(n+1-i)N_{D5}\, ,
\end{equation}
and the last gauge group is empty, in agreement with the fact that there are $k_0$ free hypermultiplets associated to the decoupled $k_0$ D5-branes.  
Each of these gauge groups has associated $M_j$ hypermultiplets in the fundamental, with $M_j$ given by,
\begin{equation}
\rho: 	\qquad N=\underbrace{1+\ldots+1}_{M_1}\,+\underbrace{2+\ldots+2}_{M_2}\, +\ldots + \underbrace{(n+1)+\ldots +(n+1)}_{M_{n+1}}
\end{equation}
These can be read from eq.(\ref{rho}) in our case.
The resulting quiver is represented in Figure \ref{quiver}, and we can see that it is fully consistent with the brane configuration in Figure \ref{stacks2}. We can also check explicitly that  
\begin{equation}
M_i+N_{i-1}+N_{i+1}>2N_i,
\label{nf2nc}
\end{equation}
a condition for the quiver to flow towards a superconformal field theory in the infrared \cite{Gaiotto:2008ak}.

\begin{figure}
\centering
\includegraphics[scale=0.45]{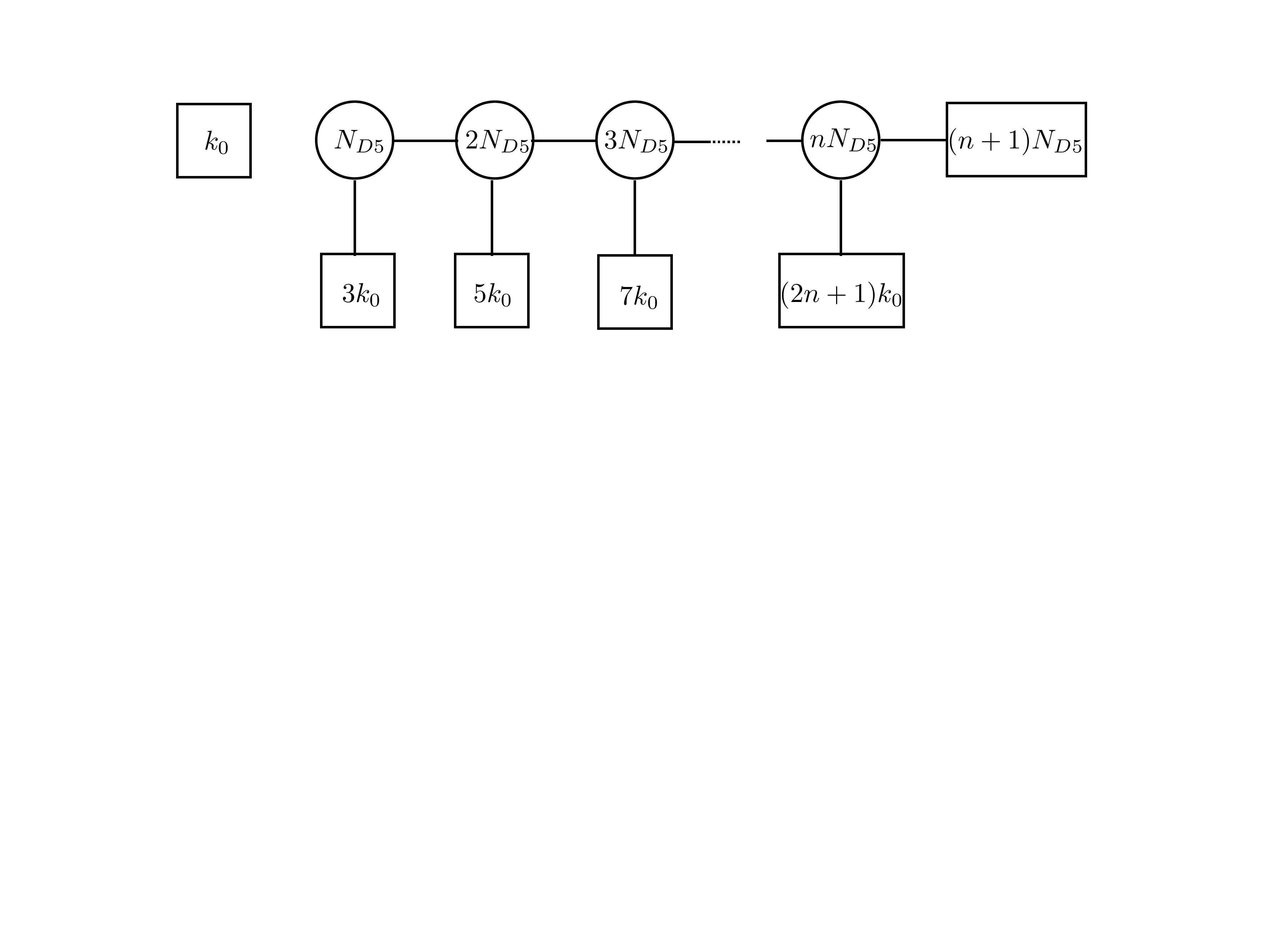}
\vspace{-7.8cm}
\caption{Quiver associated to the (NS5, D3, D5) brane set-up in Figure \ref{stacks2}.}
\label{quiver}
\end{figure} 
In summary, we have seen that {\it ending} the Hanany-Witten set-up  in Figure \ref{stacks1} and {\it completing} it with flavour branes as in Figure \ref{stacks2}, lead us to a concrete proposal
for a quiver describing a $T_\rho^{\hat{\rho}}(SU(N))$ theory. 
The charges of the non-Abelian T-dual solution calculated in eq.(\ref{eq:NATD_charges2}) were 
instrumental in identifying the quiver and its completion.
We will now show that the metric and other fields in the non-Abelian T-dual background are also consistent with those associated to the quiver of Figure \ref{stacks2}. The non-Abelian T-dual geometry will arise as a zooming-in on a particular region of the ABEG
\cite{Assel:2011xz} solution associated to the $T^{\hat \rho}_{\rho}(SU(N))$ field theory.

\section{The Type IIB $\mathcal{N}=4$ $AdS_4$ solution as a ABEG geometry}\label{seccion5bis}

 Since the solution that we generated in Section \ref{sec:sugra_solutions} preserves ${\cal N}=4 $ SUSY, contains an $AdS_4$ factor and has $SO(4)$ isometry, 
 it is natural to expect that it should fit within the formalism described in Section \ref{seccion4}. Below, we will prove this.  We start by redefining,
\begin{equation}
\sigma=-\cos\mu,\quad \beta^2 = \frac{ k^2}{L^6}.
\end{equation}
For the non-Abelian T-dual solution in eqs.(\ref{eq: IIBmetric})-(\ref{eq:NATD_Fluxes}), we can calculate,
\begin{align}
\lambda^2_1&= \frac{ r^2}{2\beta^2\Delta}(1-\sigma)(1+\sigma)^2,~~~\lambda^2_2=  \frac{(1-\sigma)^{3/2}}{\sqrt{2}\beta},~~~ \lambda^2= \frac{\sqrt{2}}{\beta}\sqrt{1-\sigma},~~~\frac{1}{\tilde{\rho}^2} = 2 \sqrt{2}\beta\sqrt{1-\sigma}(1+\sigma),\nn\\[2mm]
b_1&=\frac{r^3}{\sqrt{2}\beta \Delta}\sqrt{1-\s}(1+\s)-n\pi,~~~b_2=c_0+\frac{k}{4}\big( r^2 +  \frac{\s(\sigma^2-3)}{\beta^2}\big),~~~z =  \sigma+ i \beta r,\nn\\[2mm]
 e^{2\Phi}&=\frac{2\sqrt{2}}{k^2\beta\Delta}(1-\sigma)^{3/2},~~~\Delta=\frac{1}{\sqrt{2}\beta^3}\sqrt{1-\sigma}(1+\sigma)\big(\beta^2 r^2+\frac{(1-\sigma)(1+\sigma)^2}{2}\big),
\end{align}
where $d(c_0)=0$ and the $n \pi$ comes from the contribution to $B_2$ of $n$ large gauge transformations. 
From this we find that the functions in eqs.(\ref{eq:WN1N2})-(\ref{eq:ansatzrules}) read, 
\begin{align}
\label{h1h2NA}
N_1 &= \frac{r k^3 \Delta}{64 \sqrt{2-2\sigma}\beta^3},~~~N_2= \frac{kr(1-\sigma)}{32\beta^4},~~~W=-\frac{kr}{16\beta^2},~~~X=-\frac{k}{16\beta^3}(1+\sigma),\\[2mm]
h_1 &=\frac{kr(1+\sigma)}{4\beta},~~~h_2= \frac{1-\sigma}{2\beta},~~~h^D_1=-\frac{k(1-(2+\sigma)\s)+ (4c_0+ k r^2)\beta^2}{8 \beta^2},~~~h^D_2=\frac{1}{2}(r-n\pi).\nn
\end{align}
Notice that the functions $h_1,h_2$ are harmonic. As established in \cite{D'Hoker:2007xy}, this implies that the equations of motion of Type IIB Supergravity are satisfied.
 Also, note that  the definition of ${h^D_2}$ in each $n\pi<r<(n+1)\pi$ cell implies it is a piecewise periodic function  such that $0<h^D_2<\pi/2$.
 
 Thus, we have shown that the solution generated by non-Abelian T-duality---eqs.(\ref{eq: IIBmetric})-(\ref{eq:NATD_Fluxes}), 
 fits within the class of solutions discussed in eqs.(\ref{eq: Ansatz})-(\ref{efe}). It is worth stressing nevertheless that it does not satisfy
 the boundary conditions in eq.(\ref{bcs})---nor does it show any of the {\it isolated } singularities than can be associated to the positions $(\delta,\hat{\delta})$, of the D5 and NS5 branes.  
This suggests that the solution generated by non-Abelian T-duality could be thought of as a {\it limit} 
 of the generic solutions in eqs.(\ref{eq: Ansatz})-(\ref{efe}), along the lines of eqs.(\ref{h1h2}). We next study this in detail.

Let us start by computing the positions of the D5 and NS5 brane stacks associated to our brane configuration in Figure \ref{stacks2}. As  explained in \cite{Assel:2011xz}
and summarised in 
Section \ref{seccion4}, these positions can be computed from,
\begin{eqnarray}
l_a= \frac{2}{\pi}\sum_{b=1}^{\hat{p}} \hat{N}_5^{b} {\rm arctan}(e^{\hat{\delta}_b-\delta_a})\;\;\;\; {\hat l}_b= \frac{2}{\pi}\sum_{a=1}^p N_5^{a} {\rm arctan}(e^{\hat{\delta}_b-\delta_a}).\nonumber
\end{eqnarray}
These equations are simply solved by
\begin{equation}
e^{\hat{\delta}_b-\delta_a}=\tan{\Bigl(\frac{\pi}{2}\frac{l_a {\hat l}_b}{N}\Bigr)}.
\end{equation}
Using eqs.(\ref{linkings1}) and (\ref{linkings2}) this gives for our brane set-up
\begin{equation}
\label{deltas}
\hat{\delta}_b-\delta_a=\log{\Bigl[\tan{\Bigl(\frac{\pi}{2N}(n+1-a)(N_{D5}+k_0 b^2)\Bigr)}\Bigr]},
\end{equation}
with $N$ given by eq.(\ref{N}). 

\begin{figure}
\centering
\includegraphics[scale=0.8]{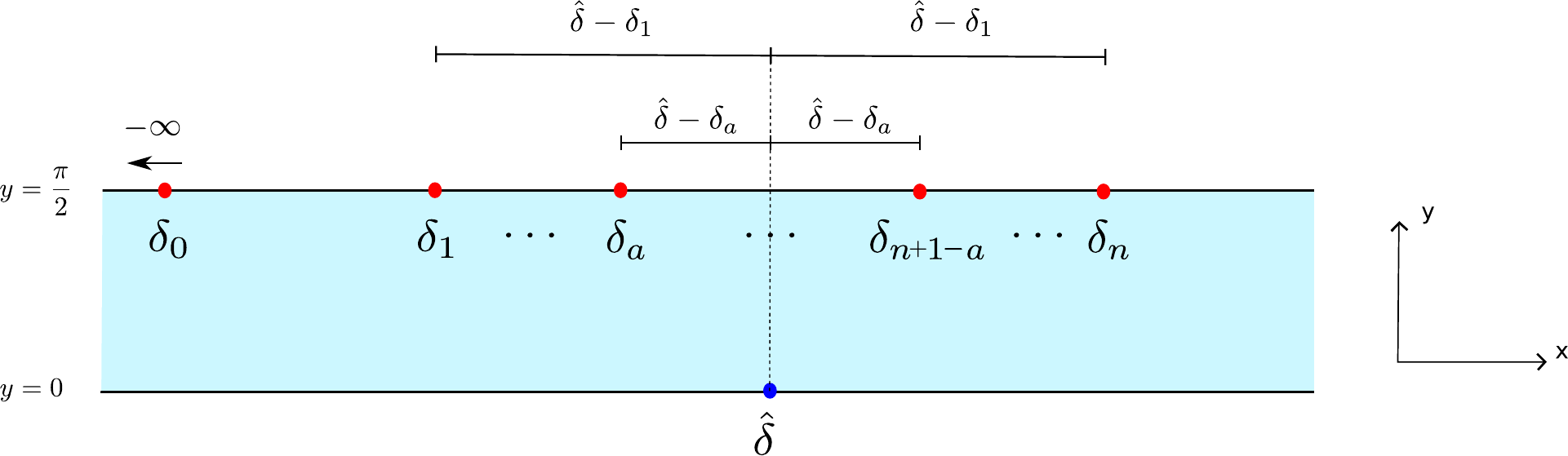}
\caption{Positions of D5 and NS5 branes in the supergravity limit. In this limit the set-up becomes symmetric around $\hat\delta=\delta_{(n+1)/2}$, with the exception of the detached stack of D5-branes at $\delta_0=-\infty$.}
\label{deltas-sugra}
\end{figure}

Recalling that we read the charges of our brane configuration from the supergravity solution, we expect to find a sensible solution to 
eq.(\ref{deltas}) in the supergravity limit  $N_{D5}\rightarrow \infty$. Taking this limit we find,
\begin{equation}
\label{deltas1}
\hat{\delta}_b-\delta_a=\log{\Bigl[\tan{\Bigl(\frac{\pi}{2}(1-\frac{a}{n+1})\Bigr)}\Bigr]}
\end{equation}
which shows that in this limit all stacks of NS5-branes can be approximately taken at the same position ${\hat \delta}$. Equivalently, we can write eq.(\ref{deltas1}) as
\begin{equation}
\label{deltas2}
\delta_a-\hat{\delta}=\log{\Bigl[\tan{\Bigl(\frac{\pi a}{2(n+1)}\Bigr)}\Bigr]}.
\end{equation}
From here we see that the first stack of (detached) $k_0$ D5-branes lies strictly at $\delta_0-{\hat \delta}=-\infty$, while the rest of stacks lie symmetrically at both sides of the NS5-branes, given that
\begin{equation}
\log{\Bigl[\tan{\Bigl(\frac{\pi c}{2(n+1)}\Bigr)}\Bigr]}=-\log{\Bigl[\tan{\Bigl(\frac{\pi a}{2(n+1)}\Bigr)}\Bigr]}  \quad {\rm for} \quad c=n+1-a .
\end{equation}
Thus,
\begin{equation}
\label{distrideltas}
 \delta_1-{\hat \delta}={\hat \delta}-\delta_n,\quad \delta_2-{\hat \delta}={\hat \delta}-\delta_{n-1},\quad \dots\quad , {\hat \delta}=\delta_{(n+1)/2}. 
\end{equation} 
This brane distribution is depicted in Figure \ref{deltas-sugra}. Let us now obtain the $h_1$, $h_2$ functions associated to this configuration, following \cite{Assel:2011xz,Assel:2012cj}.

In the supergravity limit the main contribution to $h_1$ in eq.(\ref{h's}) comes from the $n$'th stack, given that the number of branes in this stack goes with $N_{D5}$ as shown by eq.(\ref{N5a}). We can then approximate,
\begin{equation}
h_1\sim -\frac14 N_5^n \log{\tanh{(\frac{i\frac{\pi}{2}+\delta_n-z}{2})}}+cc .
\end{equation}
For $h_2$ we have in turn
\begin{equation}
h_2\sim -\frac14 (n+1)\log{\tanh{(\frac{z-{\hat \delta}}{2})}}+cc.
\end{equation}
Choosing
\begin{equation}
\delta_n=-{\hat \delta}=-\frac12 \log{[\tan{(\frac{\pi}{2(n+1)})}]},
\end{equation}
the $n$'th stack of D5-branes lies approximately at plus infinity for large $n$ while the stack of NS5-branes lies approximately at minus infinity. For finite $x$ we can then use the approximate expressions for $h_1$, $h_2$ in eq.(\ref{h1h2}) to produce,
\begin{eqnarray}
& & h_1\sim \sin{y} \,N_5^{n}\,e^{x-\delta_n}\sim\sin{y} \,N_{D5}\, \sqrt{\frac{\pi (n+1)}{2}} e^x, \nonumber\\
& & h_2\sim\cos{y}\, (n+1)\, e^{{\hat \delta}-x}\sim \cos{y}\, \sqrt{\frac{\pi (n+1)}{2}}\,  e^{-x},\label{h1h2ABEG}
\end{eqnarray}
where we have approximated $\delta_n=-{\hat \delta}\sim-\frac12 \log{[(\frac{\pi}{2(n+1)})]}$ for large $n$.
Close to $y=0$, $x=0$ we have, 
\begin{eqnarray}
\label{h1h2more}
&&h_1\sim y \,N_{D5}\, \sqrt{\frac{\pi (n+1)}{2}}\,(1+x), \nonumber\\
&&h_2\sim  \sqrt{\frac{\pi (n+1)}{2}}\, (1-x).
\end{eqnarray}

Let us now compare these expressions with those of our non-Abelian T-dual solution. Taking the functions $h_1$, $h_2$ for this solution from eq.(\ref{h1h2NA}), 
\begin{eqnarray}
\label{nonabelian}
&&h_1=\frac{kr(1+\sigma)}{4\beta},\nonumber\\
&&h_2=\frac{1-\sigma}{2\beta},
\end{eqnarray}
we find that they agree with eq.(\ref{h1h2more}) if  we identify $x= \sigma$ and
\begin{equation}
\frac{1}{2\beta}\sim  \sqrt{\frac{\pi (n+1)}{2}}\,\, , \qquad r\sim \frac{2}{k}N_{D5}\, y .
\end{equation}
Taking into account that $\beta=k/L^3$, and $N_{D5}=L^6/(\pi k)$, these are equivalent to
\begin{equation}
\label{Nd5limit}
N_{D5}\sim 2k (n+1)\,\, , \qquad r\sim 4(n+1)y .
\end{equation}
The output of this analysis is that the non-Abelian T-dual  solution comes out 
when zooming into the region $x=y=0$ of the ABEG solution associated to the brane set-up in Section \ref{seccion5}. Further, we have shown that in order to match these solutions $n$ must go to infinity as $N_{D5}/k$, which is consistent with the fact that $n$ is unbounded in the non-Abelian T-dual solution. 
Note however that this limit should be taken directly in equation (\ref{deltas}) for consistency of the previous analysis. We have checked numerically that  the matching between the non-Abelian T-dual solution and the ABEG geometry still holds in this limit in the region $x\sim 0$, $y\sim 0$.
In this matching we must have  
$\sigma\sim x$, $r\sim 4(n+1) y$. Therefore, $\sigma$, which in the non-Abelian T-dual solution ranges in $[-1,1]$, must be small.  The coordinate $r$ in turn, may cover a finite region depending on how the $y\rightarrow 0$ limit is taken in the expression above, which is unspecified in our analysis.

\begin{figure}
	\centering
	\includegraphics[scale=0.9]{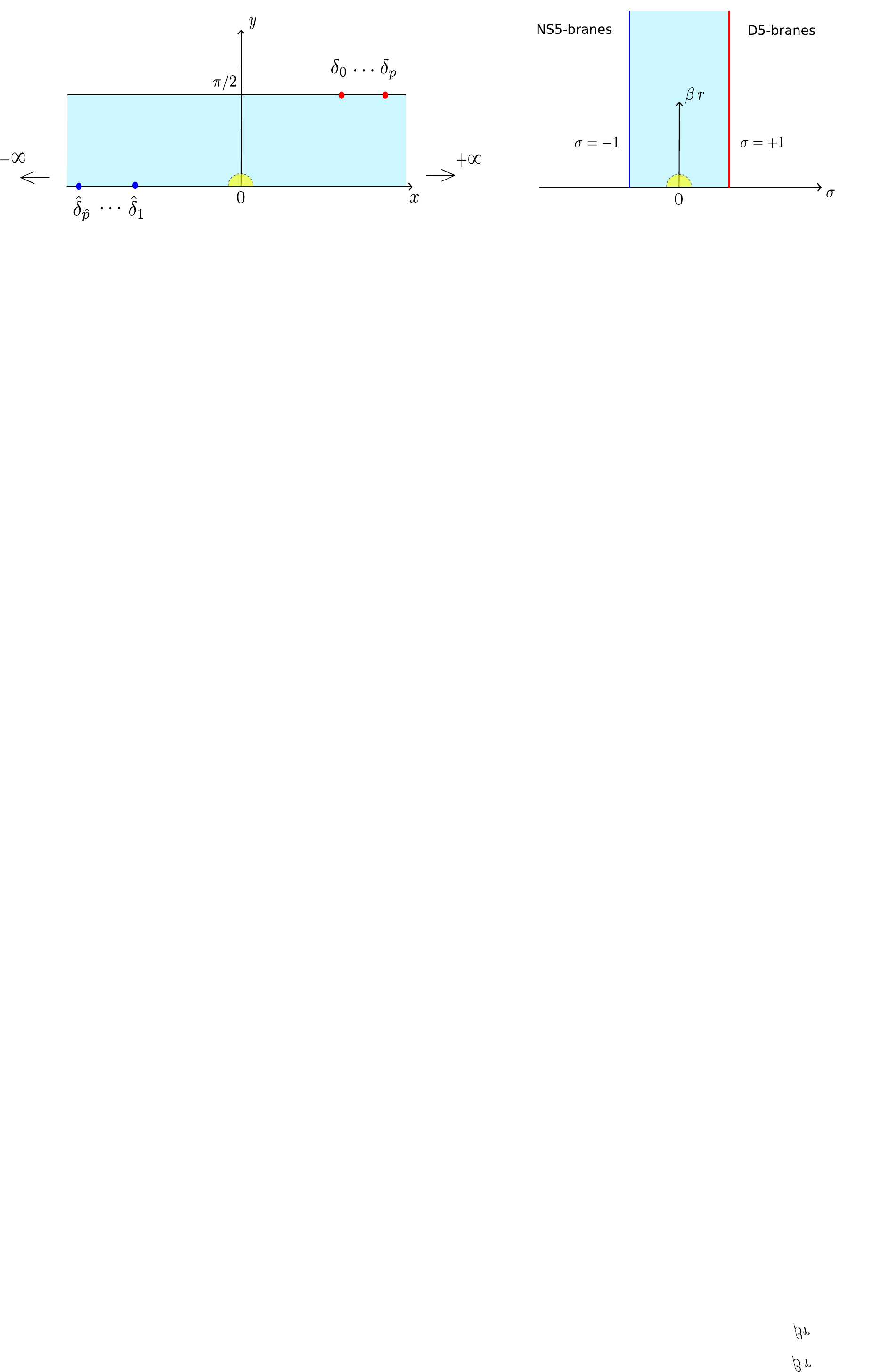}
	\vspace{-22cm}
	\caption{The ABEG set-up in the limit in which the NS5-branes are placed at $-\infty$ and the D5-branes at $+\infty$ (left). The non-Abelian T-dual set-up, with $\sigma\in[-1,1]$, $r\in[0,+\infty)$, with smeared NS5-branes at $\sigma=-1$ and smeared D5-branes at $\sigma=+1$ (right). The matching of the solutions occurs locally around $x,y\sim 0$. The \textit{completion} of the non-Abelian T-dual geometry is achieved extending $\sigma$ to $\pm\infty$ and bounding $r$ to an interval. In this completion the NS5-branes are localised at $-\infty$ and the D5-branes at $+\infty$.}
	\label{rotated_strip}
\end{figure}

The previous agreement suggests that we may see the ABEG solution as a {\it completion} of the non-Abelian T-dual geometry, that: i) Extends it to $-\infty <\sigma < \infty $, such that the singularities in $\sigma=\pm 1$ are moved to $\pm \infty$, and thus resolved, and ii) Delimits $r$ to a bounded region. This is shown pictorially in Figure \ref{rotated_strip}.
It is interesting that this completion makes explicit the ideas in \cite{Lozano:2016kum}, where a completion of the non-Abelian T-dual of $AdS_5\times S^5$ as a superposition of Maldacena-Nunez geometries was outlined.

\section{Free energy}\label{seccion6}

The authors of reference \cite{Assel:2012cp} computed the
free energy of some specific examples of $T_\rho^{\hat \rho}(SU(N))$ field theories, both directly in the field theory as well as using holography.
This free energy was shown to exhibit a $\frac12 N^2 \log{N}$ behaviour at leading order.
Further, it was argued that this value should provide an upper bound to the free energy of any, more general, $T_\rho^{\hat \rho}(SU(N))$ field theory. 

 In this section we compute the free energy associated to the non-Abelian T-dual
solution and compare it to that of the completed ABEG geometry. We show that, as expected, the free energies do not agree, consistently with the fact that the non-Abelian T-dual geometry approximates the ABEG geometry only in a small patch. On the contrary, this calculation shows explicitly how the completion of the non-Abelian T-dual geometry leads to a sensible value for the free energy of the dual  $T_\rho^{\hat \rho}(SU(N))$ field theory that is in consonance with previous results in the literature and satisfies the bound found in \cite{Assel:2012cp}.

We will use the conventions in  \cite{Assel:2012cp}. In this reference the free energy is computed from
\begin{equation}
\label{free1}
S_{\rm eff}=\frac{1}{2^4 \pi^5}{\rm Vol}_6,
\end{equation}
where ${\rm Vol}_6$ is the volume of the six dimensional internal space, which can be calculated from the functions $h_1$, $h_2$, 
defined in the 2d manifold $\Sigma_2$ as,
\begin{equation}
\label{free2}
{\rm Vol}_6=32 (4\pi)^2 \int_{\Sigma}d^2x (-W) h_1 h_2 .
\end{equation}

We first use this expression to compute the free energy associated to the non-Abelian T-dual solution. In this case we find,
\begin{equation}
h_1 h_2=\frac{k}{8\beta^2} r (1-\sigma^2)\, , \quad W=-\frac{kr}{16\beta^2},
\end{equation}
and
\begin{equation}
\label{volume}
{\rm Vol}_6=32 (4\pi)^2\frac{k^2}{2^7 \beta^4} \int_{\Sigma} r^2 (1-\sigma^2)\beta dr d\sigma .
\end{equation}
Here we have used that the differential area of the strip is $d\Sigma=\beta dr d\sigma$. Integrating  
$r\in [0,(n+1)\pi]$ and $\sigma\in [-1,1]$ we find
\begin{equation}
\label{freeenergy}
S_{\rm eff}=\frac{\pi^{3/2}}{9}\sqrt{k} N_{D5}^{3/2} (n+1)^3 .
\end{equation}
As in previous non-Abelian T-duals of $AdS$ backgrounds, this free energy exhibits the same behaviour, in this case as $\sqrt{k} N^{3/2}$, of the original $AdS$ background, multiplied by a power of
$(n+1)$  coming from the NS5-branes.

Let us now analyse the ``Abelian T-dual limit'' of this expression. This limit was first discussed in 
\cite{Lozano:2016kum} at the level of the central charges. 
 It was shown that the central charge (and the free energy) of the $SU(2)$ non-Abelian T-dual of $AdS_5\times S^5$ and that of its Abelian T-dual counterpart\footnote{Namely, the result of T-dualising the original $AdS_5\times S^5$ background along the Hopf fibre of the $S^3$ in the internal space.}
exactly match if $r$ is taken in a $r\in [n\pi,(n+1)\pi]$ interval and $n$ is sent to infinity. In this limit both metrics do in fact fully  agree. We have presented a detailed analysis of this limit in Appendix \ref{section:ATD} for the present $AdS_4$ non-Abelian T-dual solution and its Abelian T-dual counterpart (see also Appendix B). Borrowing the result for the free energy of the Abelian T-dual solution we can show that it fully agrees with the free energy of the non-Abelian solution for
$r\in [n\pi, (n+1)\pi]$ and $n\rightarrow\infty$.

Indeed, integrating $r\in [n\pi,(n+1)\pi]$, $\sigma\in [-1,1]$ in eq.(\ref{volume}) and taking the $n\rightarrow \infty$ limit, we find
\begin{equation}
\label{freeabelian}
S_{\rm eff}=\frac{\pi^{3/2}}{3}\sqrt{k} N_{D5}^{3/2} n^2 .
\end{equation}
Using now that $N_{D3}=nN_{D5}$ for the non-Abelian solution and that in the large $n$ limit $k_{D5}=nk\pi/2$, as implied by the second expression in eq.(\ref{eq:NATD_charges2}),
eq.(\ref{freeabelian}) can be rewritten as
\begin{equation}
S_{\rm eff}=\frac{\sqrt{2}\pi}{3}\sqrt{k_{D5}}N_{D3}^{3/2} .
\end{equation}
One can see that this result matches exactly the free energy of the Abelian T-dual background, given by eq.(\ref{freeenergyabelian}). As stressed in \cite{Lozano:2016kum}, this calculation shows that non-Abelian T-duality in an interval of length $\pi$ corrects the Abelian T-duality calculation by $1/n$ terms. In our present set-up this provides a non-trivial check of the validity of expression (\ref{volume}).

Let us now compare the free energy of the full non-Abelian solution, given by eq.(\ref{freeenergy}), to the free energy computed from the completed  ABEG geometry. As shown in \cite{Assel:2012cp} the approximated expressions given by (\ref{h1h2}) are enough to capture the leading order behaviour. 
Using then these approximated expressions for $h_1$, $h_2$ for our particular ABEG geometry, given by eqs.(\ref{h1h2ABEG}), we can write 
$h_1 h_2\sim \frac12 \sin{2y}  N_{D5} (n+1)^2 e^{-2\delta_n}$ and
\begin{equation}
W=\frac14 \frac{\partial^2}{\partial y^2}(h_1 h_2)=-h_1 h_2 .
\end{equation}
The internal volume then reads,
\begin{equation}
{\rm Vol}_6=16\,(4\pi)^2  N_{D5}^2 (n+1)^4 e^{-4\delta_n}\int_{0}^{\frac{\pi}{2}}dy \sin^2{2y}\int_{{\hat \delta}}^{\delta_n}dx ,
\end{equation}
which gives to leading order,
\begin{equation}
{\rm Vol}_6=16\pi^5 N_{D5}^2 (n+1)^2 \log{(n+1)},
\end{equation}
and finally,
\begin{equation}
\label{freeABEG}
S_{\rm eff}= N_{D5}^2 (n+1)^2 \log{(n+1)} .
\end{equation}
We thus see that the free energy of the $T_\rho^{\hat \rho}(N)$ theory associated to our configuration
exhibits a similar logarithmic behaviour to that of the examples discussed in \cite{Assel:2012cp}. As in those examples, the logarithm comes holographically from the size of the configuration. In our case it depends however on the number of NS5-branes, rather than on the number of D3-branes. 
Interestingly, taking into account the relation between $N$ and $n$, given by eq.(\ref{N}), the free energy given by eq.(\ref{freeABEG}) satisfies the $\frac12 N^2\log{N}$ bound suggested in  \cite{Assel:2012cp} for the free energy of general $T_\rho^{\hat \rho}(N)$ field theories. This is to our knowledge the first check in the literature of
the conjecture in \cite{Assel:2012cp}. 
 
We would like to note that for our particular $T_\rho^{\hat \rho}(N)$ theory, there is no field theoretical computation  in the literature, along the lines of  \cite{Benvenuti:2011ga,Nishioka:2011dq},  with which we could compare our holographic result. Indeed, the scaling limit taken in the field theory computation in \cite{Assel:2012cp}, given by
\begin{equation}
N_5^{a}=N^{1-\kappa_a}\gamma_a\, ; \quad l_a=N^{\kappa_a}\lambda_a ,
\end{equation}
with $0\leq \kappa_a < 1$ and $\sum_{a=1}^p \gamma_a \lambda_a = 1$, is not fulfilled by our configuration, for which only $\kappa_n=0$ is well-defined. The reason we avoid this scaling is that there is a further $N$ dependence in the number $p$ that appears in  $N=\sum_{a=1}^p N_5^{a} l_a$, as compared to the situation considered in \cite{Assel:2012cp}. It would be interesting to extend the field theory calculation in \cite{Assel:2012cp} to cover the present, more general, set-up, and check if the result matches the holographic computation.

As we have previously mentioned, we can see quite explicitly from the calculation of the free energy how the non-Abelian T-dual solution is completed by the ABEG geometry. 
Indeed, taking into account the different parametrisation of the strip in the non-Abelian T-dual solution,  $d\Sigma=\beta dr d\sigma$,  and in the ABEG solution, $d\Sigma=dx dy$,  and doing the {\it completions}
\begin{equation}
\label{extension1}
\int_{-1}^1 (1-\sigma^2)d\sigma \rightarrow \int_{\hat{\delta}}^{\delta_n}e^x e^{-x} dx,
\end{equation}
and
\begin{equation}
\label{extension2}
\beta\int_{0}^{(n+1)\pi} r^2 dr 
\rightarrow 2 (n+1)^2 \int_{0}^{\pi/2}\sin^2{2y}dy,
\end{equation}
which extend in a particular way the relations $\sigma\sim x$, $r\sim 4(n+1)y$, valid in the $x, y\sim 0$ region, we can recover exactly
the free energy associated to the ABEG solution, given by eq.(\ref{freeABEG}),  from that of the non-Abelian T-dual solution.
In this completion the logarithm is associated to the infinite extension of the configuration in the $x$ direction, which is what allows us to send the singularities in $\sigma=\pm 1$ to $\pm \infty$. Note that the completion changes as well, and quite dramatically, the $\sqrt{k} N_{D5}^{3/2} (n+1)^3$ scaling of the free energy of the non-Abelian T-dual solution into the $N_{D5}^2 (n+1)^2$ scaling associated to the ABEG geometry. Interpreting the behaviour of the free energy of $AdS$ backgrounds generated through non-Abelian T-duality has remained an interesting open problem in the non-Abelian duality literature. Indeed, in all examples analysed so far the free energy of the non-Abelian T-dual was simply that of the original background corrected by a factor of $(n+1)$ to some power, associated to the NS5-branes. One was thus led to interpret that non-Abelian T-duality was not changing too much the field theory. Instead, the detailed calculation done in the present example shows that the completion needed to correctly define the dual CFT can change this behaviour quite significantly.

To summarise, we have shown that expressions (\ref{extension1}), (\ref{extension2}) inform us about the precise way in which the non-Abelian T-dual solution must be {\it completed} in order 
to describe holographically a $T_\rho^{\hat \rho}(SU(N))$ theory:
\begin{itemize}
\item Expression (\ref{extension1}) shows that the interval $\sigma\in [-1,1]$ must be extended to $\sigma\in (-\infty,\infty)$. The two singularities at $\sigma=\mp 1$ are then  moved to infinity such that a perfectly smooth background remains. 
\item Expression (\ref{extension2}) informs us about how precisely the non-compact direction of the non-Abelian T-dual solution must be bounded.
\end{itemize}

Our $AdS_4$  example thus provides a new $AdS$ background in which the CFT dual can be used to define the geometry, in complete analogy with the $AdS_5$ case discussed in  \cite{Lozano:2016kum}. It also shows that the completion can significantly change the scaling of the free energy, and thus the CFT. This may shed some light on the possible interpretation of the behaviour of the free energy under non-Abelian T-duality.

\section{Summary and conclusions}\label{seccionconclusiones}
Let us start by summarising the contents of this paper. Then we will present some ideas for future work and comment on open problems that
our results suggest.

We started by constructing a new solution to the Type IIB equations of motion. This new background consists of an $AdS_4$  factor and two spheres $S^2_1, S^2_2$,
fibered on a Riemann surface $\Sigma(z,\bar{z})$. A dilaton, NS three form and Ramond three and five forms complete it. The system preserves sixteen supercharges and is obtained acting with non-Abelian T-duality
on the dimensional reduction of $AdS_4\times S^7$ to Type IIA. Both the original type IIA and its type IIB counterpart are singular.
An important achievement of this  paper is to understand the way of completing the geometry so that the only remaining {\it isolated} singularities are associated
with brane sources. Global aspects of the geometry have also been understood
thanks to this completion.

The procedure that we used to achieve these results can be summarised as follows. The study of the Page charges in Section \ref{sec:sugra_solutions}, suggested the brane distribution and Hanany-Witten set-up. The isometries of the background indicated the global symmetries of the dual field theory and the same goes for the amount of preserved SUSY. These data constrained our system in an important way, and suggested the way in which 
 the Hanany-Witten set up, that in principle is {\it unbounded}, can be {\it completed} (hence {\it closed}) by the addition of flavour branes. This completion, shown explicitly comparing Fig. \ref{stacks2} with Fig. \ref{stacks1}, is needed in order to define the partitions from which the $T_\rho^{\hat{\rho}}(SU(N))$ dual theory can be read.
The position (in theory space) where this completion takes place is arbitrary and determines the parameters of the dual field theory. From here, the knowledge of the associated field theory, that in this case flows to a conformal fixed point, is constraining enough to allow us to write a precise completed Type IIB background in terms of a couple of holomorphic functions defined on a Riemann surface. This background describes an intersection of D3-D5-NS5 branes, and is smooth, except at the isolated positions of the five brane sources. Then, we discussed how a particular zoom-in on a region of the  completed background gives place to the original Type IIB solution obtained 
by non-Abelian T-duality. Finally,  our calculation of the free energy showed explicitly that this completion produces a sensible result for the free energy of the associated $T_\rho^{\hat{\rho}}(SU(N))$ field theory, satisfying the upper bound $\frac12 N^2\log{N}$ found in   \cite{Assel:2012cp}. This result suggests that there could be a scaling in the field theory side that reproduces our gravitational result. 

A couple of points are crucial in the previous summary. On the one hand we have assumed that the fluxes capture faithfully the brane distribution (except, of course, for the completion with flavour branes). This has allowed us to suggest a Hanany-Witten set-up and to calculate (after completing it) the linking numbers
that select the particular CFT. On the other hand, the fact that we are using field theory knowledge to smooth out a supergravity background is quite original and key to our procedure.

Interestingly, our approach has also allowed us to find out about global properties, in particular about the range of the $r$-coordinate (which is one of the long-standing problems of the whole non-Abelian T-duality formalism). It also gives a clean way of resolving or interpreting singularities in terms of sources. This is particularly nice since the presence of these sources is a consequence of the
flavour symmetry on the field theory side, that also reflects in the completed quiver. A circle of ideas closes nicely.

{\it What remains to be done (for this particular system and more generally)?}

The proposed picture of intersecting Dp-D(p+2)-NS5 branes associated with an $AdS_{p+1}$ background should be tested in detail. For this, a more complete case-by-case study is needed. Examples with different dimensionality might reveal new subtleties, that in the present study or in that of 
\cite{Lozano:2016kum} do not show. In particular, it is clear that backgrounds  with $AdS_6$ and $AdS_3$ factors should be studied following the ideas presented here. Progress should be possible in cases with less SUSY and smaller isometry groups.

In relation to the present $AdS_4/CFT_3$ case, it would be interesting to investigate Wilson loops, vortex operators \cite{Assel:2015oxa}
and other subtle CFT aspects---
see for example \cite{Cremonesi:2014uva}, to understand, in particular, how our solutions capture these fine-points. The study of the spectrum of glueballs and mesons using our backgrounds (both the one obtained via non-Abelian T-duality and the completed one) is also of potential interest to learn about the nature of the duality. It would also be interesting to understand
the geometric realisation of the decoupled flavour group in the
quiver associated to our completed geometry

More generally, it would be very interesting to find out a precise answer for what is the effect of a non-Abelian T-duality transformation at the CFT level. In our example we started with a background dual to a CFT with one node and adjoint matter, to which we associated (after a non-Abelian T-duality transformation) a quiver containing a large number of colour and flavour groups.
But, how precisely did we go from one quiver to the other? Is an 'unhiggsing' at work, or is the non-Abelian T-duality a genuine  non-field theoretical operation? For a quite particular case of non-Abelian T-duality transformation, some progress was recently reported in
\cite{Hoare:2016wsk}.
 
Finally, it would be very nice if the ideas developed in this work could be used to answer deep questions about the nature of non-Abelian T-duality in String theory. For example,  its invertibility, or the character of the genus and $\alpha'$-expansions. We have given some evidence that the AdS/CFT correspondence can be very useful also in this regard.

\subsection*{Acknowledgements}
Some colleagues contributed with discussions and clarifications that are reflected in this paper. For these, we thank Benjamin Assel, Georgios Itsios, Diego Rodr{\'i}guez-Gomez, Daniel Thompson, Salom\'on Zacar{\'i}as.

N.T.M is supported by INFN and by the European Research Council under the European Union's Seventh Framework Program (FP/2007-2013) - ERC Grant Agreement n. 307286 (XD-STRING).
Y.L. and J.M. are partially supported by the Spanish and Regional Government Research Grants FPA2015-63667-P and
FC-15-GRUPIN-14-108. J.M. is supported by the FPI grant BES-2013-064815 of the Spanish MINECO. This work has been partially supported by the EU-COST Action MP1210, in particular through a Short Term Scientific Mission of J.M. to Swansea U. Y.L. and J.M. would like to thank the Physics Department of Swansea U. for the warm hospitality. C. Nunez is Wolfson Fellow of the Royal Society.

\begin{appendix}

\section{The Abelian T-dual limit}\label{section:ATD}

In this Appendix we summarise the key properties of the $\mathcal{N}=4$ $AdS_4$ type IIB solution that is generated from the IIA solution in equation \eqref{eq:IIA sol} by T-duality along the Hopf direction of $S^3_1$. This solution, dual to a circular quiver, was discussed at length in \cite{Assel:2012cj}. Here we recall its more relevant properties in the notation used in this paper. 
We show that it emerges as the $r\rightarrow \infty$ limit of the non-Abelian T-dual solution in Section 2 (see Appendix B). In this limit the free energies of both solutions also agree, as shown in section 6.

\subsection{The solution}

The Abelian T-dual of the IIA solution in equation \eqref{eq:IIA sol} along the Hopf direction of $S^3_1$
reads:
\begin{align}\label{eq: IIBmetab}
ds^2_{IIB} =& e^{\frac{2}{3}\phi_0}\cos\left(\frac{\mu}{2}\right)\bigg[ ds^2(AdS_4) +L^2\bigg( d\mu^2+ \frac{k^2}{ L^6 \sin^2\left(\mu\right)}dr^2+  \cos^2\left(\frac{\mu}{2}\right)ds^2(S^2_2)\bigg)\bigg]\nn\\
+&\frac{L^3}{k}\sin\left(\frac{\mu}{2}\right)\sin(\mu)ds^2(S^2_1),~~~
B_2= \cos{\theta_1}d\phi_1\wedge dr,~~~  e^{2\Phi}= \frac{e^{4\phi_0/3}}{L^2 \tan^2\left(\frac{\mu}{2}\right)}.
\end{align}
Since we dualise on the Hopf fibre $0<\psi_1<4\pi$, we have $0<\tilde{\psi}_1<\pi$ \footnote{Recall that the periodicity of the Abelian T-dual coordinate is fixed by the condition 
$\int d\psi_1\wedge d{\tilde \psi}_1=(2\pi)^2$.}. To ease notation we choose to label $\tilde{\psi_1}=r$, as we have a similar coordinate in the non-Abelian T-dual case of Section \ref{sec:sugra_solutions}, here though we stress that it is compact.
Additionally this background is supported by the gauge invariant RR fluxes
\beq\label{eq:ATD_RR}
F_3=\frac{k}{2} dr\wedge\text{Vol}(S^2_2),~~~
F_5= -\frac{3}{L} dr\wedge \text{Vol}(AdS_4) +\frac{3 L^6}{4k}\sin^3(\mu) d\mu\wedge\text{Vol}(S^1_2)\wedge \text{Vol}(S^2_2).
\eeq
As observed in \cite{Macpherson:2015tka,Lozano:2016kum} for other $AdS$ backgrounds (see Appendix B for a general analysis), this solution arises in the $r\rightarrow \infty$ limit of the non-Abelian T-dual solution derived in  Section \ref{sec:sugra_solutions}. This is straightforward for the metric and the NS-NS 2-form\footnote{Note that $B_2$ arises in the gauge $B_2=r {\rm Vol}(S^2_1)$ in the Abelian T-dual.}, while  
the dilatons differ by an $r^2$ factor that accounts for the different integration measures in the partition functions of the Abelian and non-Abelian T-dual $\sigma$-models, as explained in \cite{Lozano:2016kum}. The RR sector, even if the fields are different (see Appendix B), yields to the same quantised charges, as we show below. Finally, in order to match both solutions globally, $r$ must live in an interval of length $\pi$, $r\in [n\pi, (n+1)\pi]$, with $n\rightarrow\infty$. It is indeed in this limit in which there is precise agreement between the corresponding free energies.

The Abelian T-dual solution is also $\mathcal{N}=4$ supersymmetric, as discussed in section 2,
and has two singularities. The first singularity at $\mu=\pi$ is inherited from the stack of D6 branes in IIA. Indeed, one finds that for $\mu\sim \pi$
\beq\label{eq:ATD_D5_smeared}
ds^2 \sim \frac{e^{2\phi_0/3}}{2}\bigg[\sqrt\nu\bigg(ds^2(AdS_4)+L^2 ds^2(S^2_1)\bigg)+ \frac{L^2}{4\sqrt{\nu}}\bigg(d\tilde{r}^2+d\nu^2+\nu^2 ds^2(S^2_2)\bigg)\Bigg],~~~e^{\Phi}\sim \frac{e^{2\phi_0/3}}{2L}\sqrt{\nu}
\eeq
which is the metric close to flat space smeared D5's, where we have defined $\tilde{r}=4 e^{-2\phi_0/3}/L^2 r$ and $\nu=(\pi-\mu)^2$.
The second singularity at $\mu=0$ is caused by NS5 branes localised there, wrapping $S^2_2$ and smeared along $r$. This is a generic result of T-dualising on the Hopf fibre of a 3-sphere with vanishing radius. Close to $\mu=0$ one finds the metric (now $\nu=\mu^2$)
\beq\label{eq:ATD_NS5_smeared}
ds^2 \sim e^{2\phi_0/3}\bigg[ds^2(AdS_4)+ L^2 ds^2(S^2_2)+ \frac{L^2}{4\nu}\bigg(d\tilde{r}^2+d\nu^2+\nu^2 ds^2(S^2_1)\bigg)\bigg],~~~e^{\Phi}\sim \frac{2e^{2\phi_0/3}}{L\sqrt{\nu}},
\eeq
as expected.

The Page charges of this solution are given by
\begin{align}\label{eq:ATD_charges}
N_{D3} &= \frac{1}{2\kappa_{10}^2 T_3}\int_{\Xi_2}\left( F_5-B_2\wedge F_3\right)= \frac{N_{D2}}{2},\nn\\[2mm]
k_{D5} &= \frac{1}{2\kappa_{10}^2 T_5}\int F_3=  \frac{k}{2} ,\nn\\[2mm]
N_{NS5}&= \frac{1}{2\kappa_{10}^2 T_{NS5}}\int_{(r,S^2_1)}H_3= 1,
\end{align}
where $\Xi_1= (r, S^1_2,S^2_2)$,~ $\Xi_2= (\mu, S^1_2,S^2_2)$ and we keep $L$ defined as it was for IIA in eq \eqref{eq: Lrule1}. 
The factors of 2 in $k_{D5}$, $N_{D3}$ originate from the different periodicities of the original and T-dual variables. They are usually absorbed through a redefinition of Newton's constant. Comparing these charges with those of the non-Abelian T-dual solution, given in expressions 
(\ref{eq:NATD_charges}), (\ref{eq:NATD_charges2}), we find that
\begin{equation}
N_{D3}^{NATD}=n\pi N_{D3}^{ATD}\, , \qquad k_{D5}^{NATD}=n\pi k_{D5}^{ATD}
\end{equation}
This same rescaling was found in \cite{Lozano:2016kum} in the matching between the Abelian and non-Abelian T-dual $AdS_5$ spaces studied in that paper. As discussed there, the $n\pi$ factor can again be safely absorbed through a redefinition of Newton's constant. We give further details of the general relationship between T-dual and non-Abelian T-dual solutions in section \ref{sec:B}.

\begin{figure}
\centering
\includegraphics[scale=0.45]{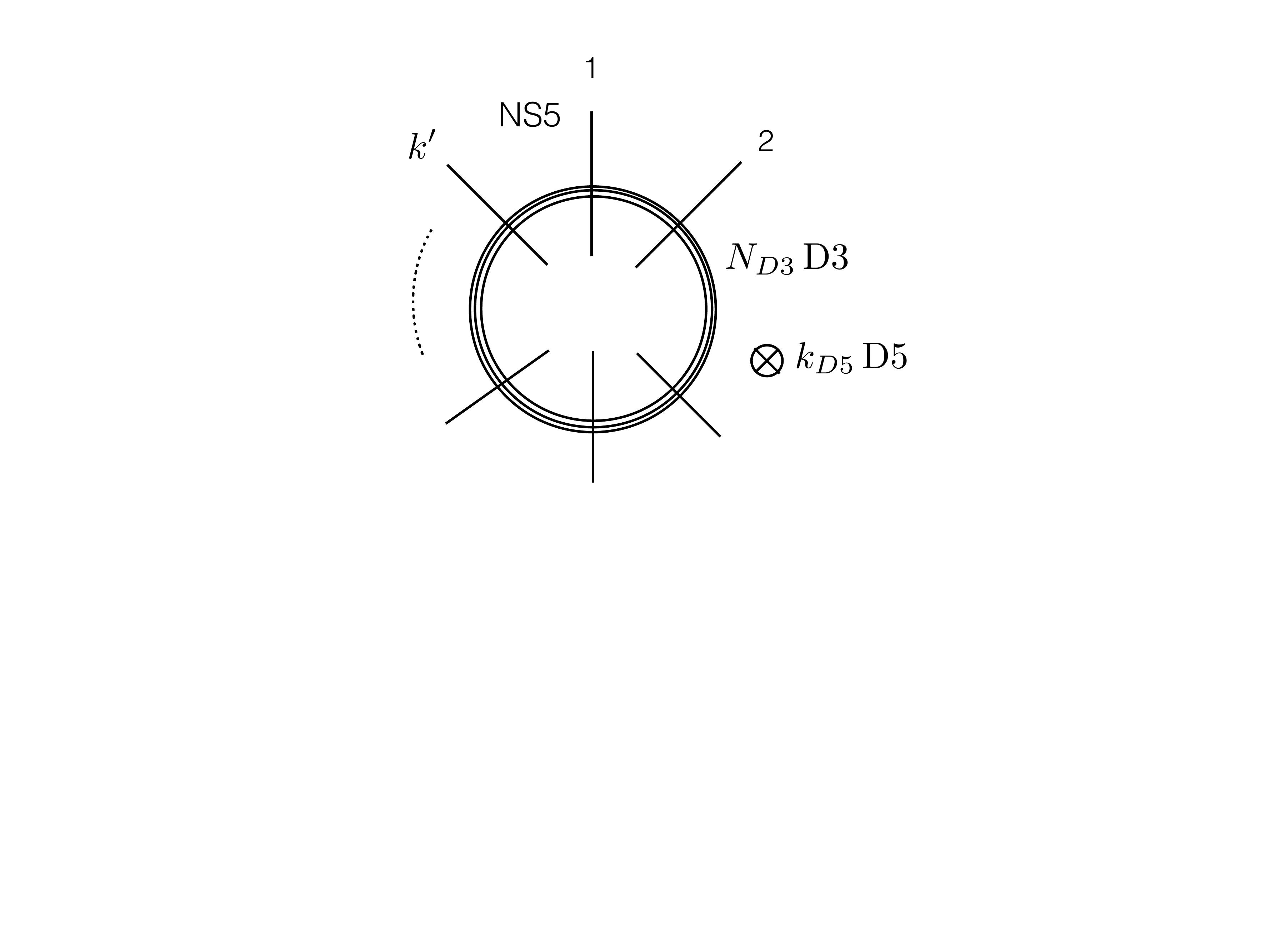}
\vspace{-6cm}
\caption{Brane set-up for the Abelian T-dual of the IIA reduction of $AdS_4\times S^7/(\mathbb{Z}_k\times \mathbb{Z}_{k'})$. At each interval there are $N_{D3}$ D3-branes stretched between the NS5-branes and $k_{D5}=k/2$ transverse D5-branes.}
\label{circular}
\end{figure}

A simple generalisation is to allow  $0<r<k'\pi$, 
which is equivalent to taking the T-dual of the IIA reduction of the $AdS_4\times S^7/(\mathbb{Z}_k\times \mathbb{Z}_{k'})$ orbifold. In that case $N_{NS5}=k^\prime$.
The solution described in \cite{Assel:2012cj} corresponds to this situation. Its CFT dual consists of a circular quiver associated to a set of $N_{D3}$ D3-branes, with $N_{D3}$ as in (\ref{eq:ATD_charges}), stretched between $k^\prime$ NS5-branes, as illustrated in Fig. \ref{circular}. These D3-branes are thus winding D3-branes. At each interval of length $\pi$ there are also $k_{D5}$ D5-branes.  The field theory associated to this brane configuration was studied in \cite{Assel:2012cj} and denoted as 
$C^{\hat \rho}_\rho (SU(N),L)$, with the positive integer $L$ refering to the number of winding D3-branes. These theories degenerate to the $T^{\hat \rho}_\rho(SU(N))$ theories of \cite{Gaiotto:2008ak} when $L=0$. In the next subsection we illustrate the connection between the 
solution in \cite{Assel:2012cj}  for 
$k^\prime=1$ and the Abelian T-dual solution under discussion. The value $k^\prime=1$ corresponds to the limiting case of $N_{D3}$ D3-branes stretched between two NS5-branes that are identified.

\subsection{Connection with ABEG geometries}

As in the case of the non-Abelian T-dual solution, we expect that the Abelian T-dual, which also preserves 
$\mathcal{N}=4$ SUSY, contains an $AdS_4$ factor and has $SO(4)$ isometry, fits within the formalism described in section 3.2. Indeed, from eqs (\ref{eq: IIBmetab},\ref{eq:ATD_RR}) we can read off the values
\begin{align}
\lambda^2_1&=\frac{\sqrt{(1-\sigma)}(1+\sigma)}{\sqrt{2}\beta} \,~~~\lambda^2_2=\frac{(1-\sigma)^{3/2}}{\sqrt{2}\beta}  \,~~~ \lambda^2=\frac{\sqrt{2}\sqrt{1-\sigma}}{\beta} ,~~~\frac{1}{\tilde{\rho}^2} = 2 \sqrt{2}\beta\sqrt{1-\sigma}(1+\sigma),\nn\\[2mm]
b_1&=r-n\pi,~~~b_2=\frac{k}{2} r+ c_0,~~~z =  \sigma+ i \beta r,~~~ e^{2\Phi}=4\frac{1-\sigma}{k^2(1+\sigma)},
\end{align}
where $d(c_0)=0$ and $\beta$ and $\sigma$ are defined as in section 5. In terms of the classification we find that the T-dual solution is given by
\begin{align}
N_1 &=\frac{k^3}{128\beta^4}(1+\sigma),~~~N_2=\frac{k}{32\beta^4}(1-\sigma) ,~~~W=-\frac{k}{16\beta^2},~~~X=0,\\[2mm]
h_1 &=\frac{k}{4\beta}(1+\sigma),~~~h_2=\frac{1}{2\beta}(1-\sigma),~~~h^D_1=-\frac{1}{2}(c_0+ \frac{k}{2}r),~~~h^D_2=\frac{1}{2}r.\nn
\end{align}

As discussed in \cite{Assel:2012cj}, this solution does not fit however within the Ansatz of \cite{Assel:2011xz}. Clearly, even if there are NS5 and D5 branes located at $\sigma=\mp 1$, which could be taken as boundaries of an infinite strip, the branes are smeared in $r$  by construction, and $h_1$, $h_2$ do not exhibit logarithmic singularities at the locations of the branes. The authors of 
\cite{Assel:2011xz} showed in \cite{Assel:2012cj} how to solve this problem.
They considered a distribution $(\delta_a,\hat\delta_b)$ of 5-branes that is repeated infinitely-many times along the strip with a period $2t$, such that $\delta_{a+p}-\delta_a=\hat\delta_{b+\hat p}-\hat\delta_b=2t$, where $p$ and $\hat p$ are, correspondingly, the total numbers of D5 and NS5-branes stacks. The resulting $h_1$, $h_2$ 
are, by construction, periodic under $z\to z+2t$. This allows for the identification of points separated by the period $2t$, thus turning the strip into an annulus (and thus the linear quiver into a circular quiver) in the $e^{i\pi z/t}$ plane, with NS5 (D5) brane stacks along the inner (outer) boundaries. The smearing of the branes comes out as a result of taking the limit $t\to0$ combined with a far-from-the-boundaries approximation. $h_1$ and $h_2$ become then independent of $r$ and non-singular. 
  
The introduction of the period $2t$ on the gravity side induces the winding D3-branes on the dual quiver, in a way that we specify below. These branes do not end on the 5-branes and therefore do not contribute to the linking numbers. The corresponding circular quiver is then characterised by two partitions $\rho$ and $\hat{\rho}$, together with the number of winding D3-branes. 
The $t\to0$ limit that yields the Abelian T-dual solution corresponds to a large number of these winding D3's, which are then identified with the $N_{D3}$ in \eqref{eq:ATD_charges}. In this approximation the number of D3-branes ending on 5-branes is negligible, and the brane picture depicted in Fig. \ref{circular} arises\footnote{Indeed, in this approximation the period $2t$ is simply related to $N_{D3}$ as $N_{D3}=(\pi^2 k k^\prime)/(32 t^2)$.}.
  
For the sake of transparency, let us finally show that the IIB NS-sector derived in \cite{Assel:2012cj} for $k^\prime=1$ NS5-branes matches our Abelian T-dual solution \footnote{Up to a scaling factor and an S-duality transformation, given e.g. in (6.1) of \cite{Assel:2012cj} for $c=b=0$ and $a=d=-1$. Note also that  \cite{Assel:2012cj}  uses a non-standard form of the dilaton, $\phi' \equiv \Phi/2$.}. The Einstein frame metric and dilaton in  \cite{Assel:2012cj}  are
\begin{align} \label{eq:Gomis_metric_IIB_ATD}
  ds^2_{IIB}&=R^2 g(y)^{1/4}\left[ ds^2_{AdS_4}  + y ds_{S_1^2}^2 + (1-y) ds_{S_2^2}^2\right] + R^2 g(y)^{-3/4} \left[ \frac{4t^2}{\pi^4}dx^2 +dy^2 \right] \;,\nonumber\\
   e^{2 \phi'}&= \frac{k^\prime}{k}\sqrt{\frac{1-y}{y}} \;,
\end{align}
where the $AdS_4$ space is taken to be of unit radius, $R^4= \pi^4 k k^\prime/t^2$ and $g(y)=y(1-y)$. If expressed in string frame by multiplying with $e^{\phi'}=e^{\Phi/2}$ and allowing for the coordinate change
\begin{equation}
\label{coordinate}
 y= \sin^2\left(\frac{\mu}{2}\right) , \quad x=\frac{1}{4}r \,,
\end{equation}
the solution in equation  \eqref{eq: IIBmetab} is reproduced for $k^\prime=1$, for which  $r\in [0,\pi]$. 
It can also be easily checked that with this coordinate change the RR-sector in \eqref{eq:ATD_RR} corresponds to the one of \cite{Assel:2012cj}, up to a gauge transformation.



\subsection{Free energy}

Using the results of the previous subsection, we can compute the free energy of the Abelian  T-dual solution from $W$, $h_1$ and $h_2$ using expressions (\ref{free1}) and (\ref{free2}) (see \cite{Assel:2012cj}). 
Taking the differential area of the strip $d^2x=\beta dr d\sigma$ and integrating in $r\in [0,\pi]$, $\sigma\in [-1,1]$, we find
\begin{equation}
S_{\rm eff}=\frac{k^2}{3\pi^2\beta^3},
\end{equation}
and, using the conserved charges in  (\ref{eq:ATD_charges}),
\begin{equation}
\label{freeenergyabelian}
S_{\rm eff}= \frac{\sqrt{2}\pi}{3}\sqrt{k_{D5}}N_{D3}^{3/2}.
\end{equation}
It can easily be checked that this is the free energy of the IIA reduction of the $AdS_4\times S^7/\mathbb{Z}_k$ orbifold, with $N_{D2}\rightarrow N_{D3}$ and $k\rightarrow k_{D5}$. It is shown in the main text that it agrees with the free energy of the non-Abelian T-dual solution in the $r\in [n\pi,(n+1)\pi]$ interval and $n\rightarrow \infty$.
  
\section{Relating Abelian and non-Abelian T-duality}\label{sec:B}

In the previous Appendix we discussed the relationship between the Abelian and non-Abelian T-dual $AdS_4$ spaces studied in this paper. 
In this appendix we complete this analysis and
elucidate a general relationship between the geometries generated by acting on a round $S^3$ with Hopf fibre T-duality and $SU(2)$ non-Abelian T-duality.

Consider a type II supergravity solution with global $SO(4)$ isometry and  NS sector that can be written as
\beq
ds^2= ds^2(\mathcal{M}_7) + 4e^{2C} ds^2(S^3),~~~B=0,~~~ e^{\Phi}=e^{\Phi_0}
\eeq
where $x$ are coordinates on $\mathcal{M}_7$ only. Non-Abelian T-duality acting on such solutions was considered at length in \cite{varios1}. It will be useful to  parametrise the 3-sphere in two different ways, making manifest the two dualisation isometries
\beq
4ds^2(S^3_{U(1)})=d\theta^2+ \sin^2\theta d\phi^2+ (d\psi + \cos\theta d\phi)^2,~~~4ds^2(S^3_{SU(2)})= \big(\omega_1^2+\omega_2^2+\omega_3^3\big),
\eeq
where $\omega_i$ are $SU(2)$ left invariant 1-forms. The first of these is suitable for T-duality on $\psi$ which, following \cite{Kelekci:2014ima}, results in the dual NS sector
\begin{align}\label{eq:ATDgen}
ds^2_{ATD}&=ds^2(\mathcal{M}_7) + e^{-2C}dr^2+e^{2C} ds^2(S^2) ,\nn\\[2mm]
B^{ATD}_2&= r \text{Vol}(S^2),~~~ e^{-\Phi_{ATD}} = e^{C-\Phi_0},
\end{align}
where we have performed a gauge transformation on $B_2$ to put it in this form, and $S^2$ is the unit norm 2-sphere spanned by $\theta, \phi$. The second sphere parametrisation is suitable for $SU(2)$ non-Abelian T-duality and leads to the dual NS sector
\begin{align}\label{eq:NATDgen}
ds^2_{NATD}&=ds^2(\mathcal{M}_7) + e^{-2C}dr^2+\frac{e^{2C}r^2}{r^2+ e^{4C}} ds^2(S^2) ,\nn\\[2mm]
B^{NATD}_2&= \frac{r^3}{r^2+ e^{4C}} \text{Vol}(S^2),~~~ e^{-\Phi_{NATD}} = \sqrt{r^2+ e^{4C}}e^{C-\Phi_0}.
\end{align}
Comparing eqs \eqref{eq:ATDgen} and  \eqref{eq:NATDgen} one finds they obey the relation
\beq
\lim_{r\to\infty}ds^2_{NATD}=ds^2_{ATD},~~~\lim_{r\to\infty}B^{NATD}_2=B^{ATD}_2,~~~ \lim_{r\to\infty}e^{-\Phi_{NATD}}= r e^{-\Phi_{ATD}}.
\eeq
This has been observed in the previous appendix and before, for instance in \cite{Lozano:2016kum}, but what has not been addressed is whether such a relation holds also for the RR fluxes. We now address this by considering the massive IIA fluxes
\begin{align}
F_0&=m,\nn\\[2mm]
F_2&= G_2,\nn\\[2mm]
F_4&= G_4 + 8 G_1\wedge  \text {Vol}(S^3),
\end{align}
however the following statements also hold when transforming from type IIB to IIA.
Performing T-duality on the Hopf fibre as before leads to the dual fluxes
\begin{align}\label{eq:FluxATDgen}
F^{ATD}_1&=-m dr,\nn\\[2mm]
F^{ATD}_3&=-dr\wedge G_2- G_1\wedge \text{Vol}(S^2)\nn\\[2mm]
F^{ATD}_5&= -dr\wedge G_4 + e^{3C}\star_7 G_4\wedge \text{Vol}(S^2)
\end{align}
while performing non-Abelian T-duality on the whole $S^3$ leads to
\begin{align}\label{eq:FluxNATDgen}
F^{NATD}_1&=-G_1- m r dr,\nn\\[2mm]
F^{NATD}_3&=e^{3C}\star_7 G_4- r dr\wedge G_2- \frac{r^3}{r^2+e^{4C}}G_1\wedge \text{Vol}(S^2)+\frac{m r^2e^{4C}}{r^2+ e^{4C}}dr \wedge \text{Vol}(S^2)\nn\\[2mm]
F^{NATD}_5&=-rdr\wedge G_4+ \frac{r^2 e^{4C}}{r^2+e^{4C}}dr\wedge G_2\wedge\text{Vol}(S^2) +\frac{r^3 e^{3C}}{r^2+e^{4C}}\star_7 G_4\wedge \text{Vol}(S^2)- e^{3C}\star_7G_2
\end{align}
Comparing eqs \eqref{eq:FluxATDgen} and \eqref{eq:FluxNATDgen}, one sees that there is indeed a relation between the flux polyforms, namely
\beq
\partial_r(\lim_{r\to\infty}F^{NATD})= F^{ATD},
\eeq
which $e^{-\Phi_{NATD}}$ clearly also obeys. Notice that we can dispense with the derivative by weighting the flux polyform by the dilaton, namely
\beq\label{eq: fluxlimitnoder}
\lim_{r\to \infty}e^{\Phi_{NATD}}F^{NATD}= e^{\Phi_{ATD}}F^{ATD}
\eeq
That this holds is actually not so surprising. As shown in  \cite{Sfetsos:2010uq,Hassan:1999bv}, under T-duality the fluxes transform in the combination $e^{\Phi}F$ . Specifically the fluxes and MW Killing spinors are transformed by the same matrix $\Omega$ as
\begin{align}\label{eq:Fluxtrans}
&\epsilon_1= \epsilon^0_1,\nn\\[2mm]
&\epsilon_2=\Omega \epsilon^0_2,\nn\\[2mm]
& e^{\Phi}F=e^{\Phi^0}F^0\Omega^{-1},
\end{align}
where $0$ hat denotes the seed solution. For $SU(2)$ non-Abelian T-duality performed on a round 3-sphere there exists a frame in which
\beq
\Omega^{NATD} = \frac{1}{\sqrt{r^2+e^{4C}}} \left(\Gamma_{r12}+ r \Gamma_r\right)
\eeq
where the flat directions $1,2$ span $e^{2C}S^2$ in the non-Abelian T-dual. Clearly
\beq
\lim_{r\to\infty}\Omega^{NATD}=\lim_{r\to\infty}\left(\Omega^{NATD}\right)^{-1}= \Gamma_r
\eeq
which we recognise as $\Omega^{ATD}$. This means that
\beq
\lim_{r\to\infty}e^{\Phi^0}F^0(\Omega^{NATD})^{-1} = e^{\Phi^0}F^0(\Omega^{ATD})^{-1} 
\eeq
and so eq \eqref{eq: fluxlimitnoder} just reconciles this with the final expression in eq \eqref{eq:Fluxtrans}.

To conclude, we have observed that the Hopf fibre T-dual is related to the non-Abelian T-dual as
\beq
\lim_{r\to\infty}\left(\begin{array}{c}ds^2\\[2mm]
B_2\\[2mm]
e^{\Phi}F\\[2mm]
\epsilon_{1,2}
\end{array}\right)_{NATD}=~~~~~\left(\begin{array}{c}ds^2\\[2mm]
B_2\\[2mm]
e^{\Phi}F\\[2mm]
\epsilon_{1,2}\end{array}\right)_{ATD}
\eeq
while the dilaton is related as
\beq
\lim_{r\to\infty}e^{-\Phi_{NATD}}= r e^{-\Phi_{ATD}}.
\eeq
 As discussed below eq \eqref{eq:NATD_D5_smeared}, it is easy to understand the $r$ appearing in the dilaton at the level of the string frame supergravity actions, where this factor precisely cancels the change in the volume of the T-dual submanifold in the NS sector. In the RR sector, it is the combination $e^\Phi F$ that absorbs the volume change.
The $r\to\infty$ limit of the non-Abelian T-dual thus reproduces the Abelian T-dual.

\end{appendix}


\begin{thebibliography}{99}


\bibitem{Montonen:1977sn}
  C.~Montonen and D.~I.~Olive,
  Phys.\ Lett.\ B {\bf 72}, 117 (1977).

\bibitem{Seiberg:1994rs}
  N.~Seiberg and E.~Witten,
  Nucl.\ Phys.\ B {\bf 426}, 19 (1994)
  Erratum: [Nucl.\ Phys.\ B {\bf 430}, 485 (1994)]
  [hep-th/9407087].
  N.~Seiberg and E.~Witten,
  Nucl.\ Phys.\ B {\bf 431}, 484 (1994)
  [hep-th/9408099].
\bibitem{Seiberg:1994pq}
  N.~Seiberg,
  Nucl.\ Phys.\ B {\bf 435}, 129 (1995)
  [hep-th/9411149].
\bibitem{Sen:1998kr}
  A.~Sen,
  In *Cambridge 1997, Duality and supersymmetric theories* 297-413
  [hep-th/9802051].
  N.~A.~Obers and B.~Pioline,
  Phys.\ Rept.\  {\bf 318}, 113 (1999)
  [hep-th/9809039].



\bibitem{Maldacena:1997re} 
  J.~M.~Maldacena,
  Int.\ J.\ Theor.\ Phys.\  {\bf 38}, 1113 (1999)
  [Adv.\ Theor.\ Math.\ Phys.\  {\bf 2}, 231 (1998)]
  [hep-th/9711200];
  S.~S.~Gubser, I.~R.~Klebanov and A.~M.~Polyakov,
  Phys.\ Lett.\ B {\bf 428}, 105 (1998)
  [hep-th/9802109];
  E.~Witten,
  Adv.\ Theor.\ Math.\ Phys.\  {\bf 2}, 253 (1998)
  [hep-th/9802150].






\bibitem{Kramers:1941kn}
  H.~A.~Kramers and G.~H.~Wannier,
  Phys.\ Rev.\  {\bf 60}, 252 (1941).
\bibitem{Coleman:1974bu}
  S.~R.~Coleman,
  Phys.\ Rev.\ D {\bf 11}, 2088 (1975);
  S.~Mandelstam,
  Phys.\ Rev.\ D {\bf 11}, 3026 (1975).


\bibitem{Buscher:1987qj}
  T.~H.~Buscher,
  Phys.\ Lett.\ B {\bf 201}, 466 (1988);
  T.~H.~Buscher,
  Phys.\ Lett.\ B {\bf 194}, 59 (1987).

\bibitem{Rocek:1991ps}
  M.~Rocek and E.~P.~Verlinde,
  Nucl.\ Phys.\ B {\bf 373} (1992) 630
  [hep-th/9110053].




\bibitem{delaOssa:1992vci} 
  X.~C.~de la Ossa and F.~Quevedo,
  Nucl.\ Phys.\ B {\bf 403}, 377 (1993)
  [hep-th/9210021].
 
 
 
 
\bibitem{Sfetsos:2010uq}
  K.~Sfetsos and D.~C.~Thompson,
  Nucl.\ Phys.\ B {\bf 846} (2011) 21
  [arXiv:1012.1320 [hep-th]].



\bibitem{varios1}
  Y.~Lozano, E.~O Colgain, K.~Sfetsos and D.~C.~Thompson,
  JHEP {\bf 1106} (2011) 106
  [arXiv:1104.5196 [hep-th]];
  G.~Itsios, Y.~Lozano, E.~O Colgain and K.~Sfetsos,
  JHEP {\bf 1208} (2012) 132
  [arXiv:1205.2274 [hep-th]];
  G.~Itsios, C.~Nunez, K.~Sfetsos and D.~C.~Thompson,
  Nucl.\ Phys.\ B {\bf 873}, 1 (2013)
  [arXiv:1301.6755 [hep-th]];
  Y.~Lozano, E.~Ó Colgáin, D.~Rodríguez-Gómez and K.~Sfetsos,
  Phys.\ Rev.\ Lett.\  {\bf 110}, no. 23, 231601 (2013)
  [arXiv:1212.1043 [hep-th]];
  G.~Itsios, C.~Nunez, K.~Sfetsos and D.~C.~Thompson,
  Phys.\ Lett.\ B {\bf 721}, 342 (2013)
  [arXiv:1212.4840 [hep-th]];
  J.~Jeong, O.~Kelekci and E.~O Colgain,
  JHEP {\bf 1305}, 079 (2013)
  [arXiv:1302.2105 [hep-th]].
  
\bibitem{Itsios:2012dc}
  G.~Itsios, Y.~Lozano, E.~O Colgain and K.~Sfetsos,
  JHEP {\bf 1208} (2012) 132
  [arXiv:1205.2274 [hep-th]];
  \bibitem{varios2}
  A.~Barranco, J.~Gaillard, N.~T.~Macpherson, C.~Núñez and D.~C.~Thompson,
  JHEP {\bf 1308}, 018 (2013)
  [arXiv:1305.7229 [hep-th]];
  N.~T.~Macpherson,
  JHEP {\bf 1311}, 137 (2013)
  [arXiv:1310.1609 [hep-th]];
  
  
\bibitem{Lozano:2013oma} 
  Y.~Lozano, E.~O.~Colgáin and D.~Rodríguez-Gómez,
  JHEP {\bf 1405}, 009 (2014)
  [arXiv:1311.4842 [hep-th]].


  
\bibitem{varios3} 
  J.~Gaillard, N.~T.~Macpherson, C.~Núñez and D.~C.~Thompson,
  Nucl.\ Phys.\ B {\bf 884}, 696 (2014)
  [arXiv:1312.4945 [hep-th]];
  D.~Elander, A.~F.~Faedo, C.~Hoyos, D.~Mateos and M.~Piai,
  JHEP {\bf 1405}, 003 (2014)
  [arXiv:1312.7160 [hep-th]];
  S.~Zacarías,
  Phys.\ Lett.\ B {\bf 737}, 90 (2014)
  [arXiv:1401.7618 [hep-th]];
  E.~Caceres, N.~T.~Macpherson and C.~Núñez,
  JHEP {\bf 1408}, 107 (2014)
  [arXiv:1402.3294 [hep-th]];
  P.~M.~Pradhan,
  Phys.\ Rev.\ D {\bf 90}, no. 4, 046003 (2014)
  [arXiv:1406.2152 [hep-th]];
  K.~Sfetsos and D.~C.~Thompson,
  JHEP {\bf 1411}, 006 (2014)
  [arXiv:1408.6545 [hep-th]].
  
\bibitem{Lozano:2014ata}
  Y.~Lozano and N.~T.~Macpherson,
  JHEP {\bf 1411} (2014) 115
  [arXiv:1408.0912 [hep-th]].
  
\bibitem{Kelekci:2014ima}
  Ö.~Kelekci, Y.~Lozano, N.~T.~Macpherson and E.~Ó.~Colgáin,
  Class.\ Quant.\ Grav.\  {\bf 32} (2015) no.3,  035014
  [arXiv:1409.7406 [hep-th]].
  
 
\bibitem{Macpherson:2014eza}
  N.~T.~Macpherson, C.~Nunez, L.~A.~Pando Zayas, V.~G.~J.~Rodgers and C.~A.~Whiting,
  JHEP {\bf 1502} (2015) 040
  [arXiv:1410.2650 [hep-th]].

 
 
\bibitem{varios4} 
  K.~S.~Kooner and S.~Zacarías,
  JHEP {\bf 1508}, 143 (2015)
  [arXiv:1411.7433 [hep-th]];
  T.~R.~Araujo and H.~Nastase,
  Phys.\ Rev.\ D {\bf 91}, no. 12, 126015 (2015)
  [arXiv:1503.00553 [hep-th]];
  Y.~Lozano, N.~T.~Macpherson, J.~Montero and E.~Ó.~Colgáin,
  JHEP {\bf 1508}, 121 (2015)
  [arXiv:1507.02659 [hep-th]];
  Y.~Lozano, N.~T.~Macpherson and J.~Montero,
  JHEP {\bf 1510} (2015) 004
  [arXiv:1507.02660 [hep-th]];
  T.~R.~Araujo and H.~Nastase,
  JHEP {\bf 1511}, 203 (2015)
  [arXiv:1508.06568 [hep-th]];
  L.~A.~P.~Zayas, V.~G.~J.~Rodgers and C.~A.~Whiting,
  JHEP {\bf 1602}, 061 (2016)
  [arXiv:1511.05991 [hep-th]].

 
\bibitem{Macpherson:2015tka} 
  N.~T.~Macpherson, C.~Nunez, D.~C.~Thompson and S.~Zacarias,
  JHEP {\bf 1511}, 212 (2015)
  [arXiv:1509.04286 [hep-th]].
 


\bibitem{Lozano:2016kum} 
  Y.~Lozano and C.~Núñez,
  JHEP {\bf 1605}, 107 (2016)
  [arXiv:1603.04440 [hep-th]].




\bibitem{Apruzzi:2014qva}
  F.~Apruzzi, M.~Fazzi, A.~Passias, D.~Rosa and A.~Tomasiello,
  JHEP {\bf 1411} (2014) 099
   Erratum: [JHEP {\bf 1505} (2015) 012]
  [arXiv:1406.0852 [hep-th]].
  
\bibitem{Kim:2015hya}
  H.~Kim, N.~Kim and M.~Suh,
  Eur.\ Phys.\ J.\ C {\bf 75} (2015) no.10,  484
  [arXiv:1506.05480 [hep-th]].
  
\bibitem{Kelekci:2016uqv}
  Ö.~Kelekci, Y.~Lozano, J.~Montero, E.~Ó.~Colgáin and M.~Park,
  Phys.\ Rev.\ D {\bf 93} (2016) no.8,  086010
  [arXiv:1602.02802 [hep-th]].
\bibitem{D'Hoker:2016rdq}
  E.~D'Hoker, M.~Gutperle, A.~Karch and C.~F.~Uhlemann,
  JHEP {\bf 1608} (2016) 046
  [arXiv:1606.01254 [hep-th]].
\bibitem{Couzens:2016iot}
  C.~Couzens,
  arXiv:1609.05039 [hep-th].












 
  

  
\bibitem{Alvarez:1993qi}
  E.~Alvarez, L.~Alvarez-Gaume, J.~L.~F.~Barbon and Y.~Lozano,
  Nucl.\ Phys.\ B {\bf 415} (1994) 71
  [hep-th/9309039];
  E.~Alvarez, L.~Alvarez-Gaume and Y.~Lozano,
  Nucl.\ Phys.\ B {\bf 424}, 155 (1994)
  [hep-th/9403155];
  Y.~Lozano,
  Phys.\ Lett.\ B {\bf 355} (1995) 165
  [hep-th/9503045].

  



\bibitem{Giveon:1993ai} 
  A.~Giveon and M.~Rocek,
  Nucl.\ Phys.\ B {\bf 421}, 173 (1994)
  [hep-th/9308154];
  S.~Elitzur, A.~Giveon, E.~Rabinovici, A.~Schwimmer and G.~Veneziano,
  Nucl.\ Phys.\ B {\bf 435}, 147 (1995)
  [hep-th/9409011].
  
  %
\bibitem{Klimcik:1995ux} 
  C.~Klimcik and P.~Severa,
  Phys.\ Lett.\ B {\bf 351}, 455 (1995)
  [hep-th/9502122].
  




















  
  
   %
\bibitem{Gaiotto:2014lca} 
  D.~Gaiotto and A.~Tomasiello,
  JHEP {\bf 1412}, 003 (2014)
  [arXiv:1404.0711 [hep-th]];
  S.~Cremonesi and A.~Tomasiello,
  arXiv:1512.02225 [hep-th].

	
\bibitem{Assel:2011xz} 
  B.~Assel, C.~Bachas, J.~Estes and J.~Gomis,
  JHEP {\bf 1108}, 087 (2011)
  [arXiv:1106.4253 [hep-th]].
\bibitem{Assel:2012cj}
  B.~Assel, C.~Bachas, J.~Estes and J.~Gomis,
  JHEP {\bf 1212} (2012) 044
  [arXiv:1210.2590 [hep-th]].
  
  
\bibitem{Aharony:2011yc} 
  O.~Aharony, L.~Berdichevsky, M.~Berkooz and I.~Shamir,
  Phys.\ Rev.\ D {\bf 84}, 126003 (2011)
  [arXiv:1106.1870 [hep-th]].





















\bibitem{D'Hoker:2007xz} 
  E.~D'Hoker, J.~Estes and M.~Gutperle,
  JHEP {\bf 0706}, 022 (2007)
  [arXiv:0705.0024 [hep-th]].
  
\bibitem{D'Hoker:2007xy} 
  E.~D'Hoker, J.~Estes and M.~Gutperle,
  JHEP {\bf 0706}, 021 (2007)
  [arXiv:0705.0022 [hep-th]].






\bibitem{Gaiotto:2008ak}
  D.~Gaiotto and E.~Witten,
  Adv.\ Theor.\ Math.\ Phys.\  {\bf 13} (2009) no.3,  721
  [arXiv:0807.3720 [hep-th]].
  
  
   
\bibitem{Hanany:1996ie} 
  A.~Hanany and E.~Witten,
  Nucl.\ Phys.\ B {\bf 492}, 152 (1997)
  [hep-th/9611230].



\bibitem{Gaiotto:2009gz} 
  D.~Gaiotto and J.~Maldacena,
  JHEP {\bf 1210}, 189 (2012)
  [arXiv:0904.4466 [hep-th]].




\bibitem{ReidEdwards:2010qs}
  R.~A.~Reid-Edwards and B.~Stefanski, jr.,
  Nucl.\ Phys.\ B {\bf 849} (2011) 549
  [arXiv:1011.0216 [hep-th]];
  O.~Aharony, L.~Berdichevsky and M.~Berkooz,
  JHEP {\bf 1208}, 131 (2012)
  [arXiv:1206.5916 [hep-th]].












  
  
  
  
  











 


 
 
\bibitem{Hassan:1999bv} 
  S.~F.~Hassan,
  Nucl.\ Phys.\ B {\bf 568}, 145 (2000)
  [hep-th/9907152].
 
 
 
 

 
 
 





 
 
 
 
 
 
 
 
 
 
 
   
  

  
  
  
\bibitem{Assel:2012cp}
  B.~Assel, J.~Estes and M.~Yamazaki,
  JHEP {\bf 1209} (2012) 074
  [arXiv:1206.2920 [hep-th]].
	
\bibitem{Benvenuti:2011ga}
  S.~Benvenuti and S.~Pasquetti,
  JHEP {\bf 1205} (2012) 099
  [arXiv:1105.2551 [hep-th]].

\bibitem{Nishioka:2011dq}
  T.~Nishioka, Y.~Tachikawa and M.~Yamazaki,
  JHEP {\bf 1108} (2011) 003
  [arXiv:1105.4390 [hep-th]].
  
\bibitem{Assel:2015oxa}
  B.~Assel and J.~Gomis,
  JHEP {\bf 1511} (2015) 055
  [arXiv:1506.01718 [hep-th]].
	
	
\bibitem{Cremonesi:2014uva} 
  S.~Cremonesi, A.~Hanany, N.~Mekareeya and A.~Zaffaroni,
  JHEP {\bf 1501}, 150 (2015)
  [arXiv:1410.1548 [hep-th]];
  S.~Cremonesi, A.~Hanany and A.~Zaffaroni,
  JHEP {\bf 1401}, 005 (2014)
  [arXiv:1309.2657 [hep-th]];
  A.~Hanany and N.~Mekareeya,
  JHEP {\bf 1201}, 079 (2012)
  [arXiv:1110.6203 [hep-th]].
	
	
	\bibitem{Hoare:2016wsk} 
  B.~Hoare and A.~A.~Tseytlin,
  arXiv:1609.02550 [hep-th].

  
  

  
  
  
  


  
  
  
  
  
  \end{thebibliography}
  \end{document}